\documentclass[journal]{IEEEtran}
\usepackage[utf8]{inputenc}
\usepackage{amsmath}
\usepackage{pdfpages}
\usepackage{amsfonts}
\usepackage{amssymb}
\usepackage{mathtools}
\usepackage{amsthm}
\usepackage{afterpage}
\usepackage{verbatim}
\usepackage{stfloats}
\usepackage{cite}
\usepackage{graphicx}
\usepackage{bm}
\usepackage[caption=false, font=footnotesize]{subfig}
\usepackage{xcolor}

\newtheorem{theorem}{Theorem}
\newtheorem{lemma}{Lemma}
\newtheorem{proposition}{Proposition}
\newtheorem{corollary}{Corollary}

\newcommand{\E}{\operatorname{E}}
\newcommand{\Var}{\operatorname{Var}}
\newcommand{\Cov}{\operatorname{Cov}}
\newcommand{\pr}{\operatorname{P}}

\begin{document}

\title{A Framework for Characterising the Value of Information in Hidden Markov Models}

\author{Zijing~Wang,~\IEEEmembership{Student~Member,~IEEE,}
        Mihai-Alin~Badiu, and~Justin~P.~Coon,~\IEEEmembership{Senior~Member,~IEEE}
\thanks{This material is based upon work supported by, or in part by, the U. S. Army Research Laboratory and the U. S. Army Research Office under contract/grant number W911NF-19-1-0048.  This work was also supported by EPSRC grant number EP/T02612X/1. The authors also gratefully acknowledge the support of the Clarendon Fund Scholarships at the University of Oxford. This paper was presented in part at the IEEE Global Communications Conference (GLOBECOM), Dec. 7-11, 2020.}
\thanks{The authors are with the Department of Engineering Science, University of Oxford, Oxford OX1 3PJ, U. K. (e-mail: \{zijing.wang, mihai.badiu, and justin.coon\}@eng.ox.ac.uk).}}

\maketitle
\begin{abstract}
In this paper, a general framework is formalised to characterise the value of information (VoI) in hidden Markov models. Specifically, the VoI is defined as the mutual information between the current, unobserved status at the source and a sequence of observed measurements at the receiver, which can be interpreted as the reduction in the uncertainty of the current status given that we have noisy past observations of a hidden Markov process. We explore the VoI in the context of the noisy Ornstein-Uhlenbeck process and derive its closed-form expressions. Moreover, we investigate the effect of different sampling policies on VoI, deriving simplified expressions in different noise regimes and analysing statistical properties of the VoI in the worst case. We also study the optimal sampling policy to maximise the average information value under the sampling rate constraint. In simulations, the validity of theoretical results is verified, and the performance of VoI in Markov and hidden Markov models is also analysed. Numerical results further illustrate that the proposed VoI framework can support timely transmission in status update systems, and it can also capture the correlation properties of the underlying random process and the noise in the transmission environment.
\end{abstract}

\section{Introduction}
\IEEEPARstart{W}{ith} the wide development of emerging 5G technologies, timely status updates are more and more important to enable real-time monitoring and control in a variety of applications, such as environmental surveillance, smart transport, industrial control, e-health, and so on. These applications can be modelled as a generic status update system in which sensor nodes are largely deployed to monitor different types of physical processes, and continuously sample data to get timely status updates about the targeted process. Stale data can be problematic. Therefore, the freshness of data plays an important role in such systems.

The age of information (AoI) was introduced in ~\cite{2012infocom?,FirstWorkMultiSource} as a new performance metric to characterise the data freshness from the receiver's perspective. It is defined as the time elapsed since the latest received status update was sampled. Specifically, the AoI at time $t$ is given as
\begin{equation}
\label{eq:AoI-definition}
    \Delta(t)=t-u(t),
\end{equation}
where $u(t)$ is the generation time of the latest sample received at the destination before time $t$. The AoI has received much attention due to its utility in characterising the timeliness of information, and it has been widely studied as a concept, a metric, and a tool in a variety of communication systems \cite{BookConceptTool,MagazineIoT}. Many works focused on AoI and its variants in different queueing systems, studying statistical properties~\cite{2012infocom?,estimationMultiSource,DistributionISITJournal}, and exploring the impact of queueing disciplines~\cite{Queue2012,QueueLCFS,QueueLCFSmultihop}, transmission priority~\cite{queuePriority}, packet deadlines~\cite{queuepacketddl}, buffer sizes and packet replacement~\cite{QueueBufferMilcom} on the performance of AoI. 

In addition to fundamental research, AoI-oriented scheduling and optimisation problems have also been studied extensively in the design of freshness-aware applications. Optimal sampling policies to minimise the AoI were formulated as Markov decision process (MDP) problems which were studied in~\cite{OptimalPolicyCDC2018,OptimalPolicyALLERTON2019,OptimalPolicyISIT2020}. In~\cite{PartialUpates}, the authors proposed that partial updates of initial samples can be used to minimise the information age. Optimal link activation and scheduling problems for AoI minimisation were investigated in single-hop~\cite{scheduleMinMaxJ} and multiple-hop networks~\cite{mineWCL}. AoI-based scheduling policies were proposed in~\cite{EHBatterySize,EHAoIscheduling} to improve energy efficiency in energy harvesting networks. The joint optimisation of trajectory design and user scheduling problems were explored in unmanned aerial vehicle (UAV) networks~\cite{UAVavergeAgeTVT,mineUAV}.  Furthermore, machine learning-based algorithms were applied to solve the above age-optimal problems more efficiently~\cite{UAVdeeplearning,UAVPacketLossQLearning,UAVlearning}. 

The age given in~\eqref{eq:AoI-definition} increases linearly with time until a new status update is received, which means that the concept of AoI is independent of the statistical variations inherent in the underlying source data. However, in some practical cases, old information may still have value while new information may have less value. For example, some information (e.g., node mobility) evolves very frequently over time; thus even fresh samples may hold little valuable information. Other information (e.g., temperature) evolves slowly; thus old samples may be sufficient enough to be used for further analysis and actuation. This means that the age of information cannot fully capture the performance degradation in information quality caused by the time lapse between status updates or the correlation properties the underlying random process might exhibit. In this regard, it seems that AoI may not be a perfect metric. Therefore, more systematic approaches should be further investigated to quantify the information value.

A general way to measure the information value is to utilise non-linear AoI functions~\cite{NonlinearSurveySY}. The authors in~\cite{QueueSYUpdatWait} proposed the concept of the ``age penalty'', which maps the AoI to a non-linear and non-decreasing penalty function to evaluate the level of ``dissatisfaction'' related to the outdated information. Closed-form expressions of non-linear age under different queueing models were derived in~\cite{NonlinearEH} for energy harvesting networks. The authors in~\cite{NonlinearVoUDJ} considered the auto-correlation of the random process and investigated exponential and logarithmic AoI penalty functions. A binary function was used to evaluate the freshness for web crawling~\cite{binaryWebCrawlers2003,binaryWebCrawlers2018} and cache updating systems~\cite{binaryCache2021}. This function takes the value $1$ when the data is up-to-date; otherwise, it takes the value $0$. This metric is appropriate for a data source that does not change very frequently. As an alternative approach, the discrete concept of age of version (AoV) was introduced to measure the difference between the version of a node and the freshest information at the source in ~\cite{AoVcache,AoIGossip,AoIGossipTopology}.

Furthermore, information-theoretic AoI research has also been widely discussed to provide a theoretical interpretation of non-linear age functions. The mean square error (MSE) in remote estimation can remove the linearity and has been extensively utilised to measure the information value~\cite{estimationMarkovSource,estimationWienerProcess,estimationOUprocess,estimationContext-aware,estimation+correlation,estimationAllerton}. In~\cite{estimationMarkovSource}, the authors defined a metric called the ``effective age'' which increases with the estimation error, and studied the optimal scheduling problem with the aim of minimising MSE for remote estimation of the Markov data source. The relationship between AoI and the estimation error was explored in the context of two Markov processes: the Wiener process~\cite{estimationWienerProcess} and the Ornstein-Uhlenbeck (OU) process~\cite{estimationOUprocess}. The authors proved that the optimal sampling policy to minimise the MSE is a threshold-based strategy. In~\cite{estimationContext-aware}, the authors defined a context-aware metric called the ``urgency of information'', which can be used to describe both the non-linear performance degradation and the context dependence of the  Markov status update system. The timely updating strategy for two correlated information sources was investigated in~\cite{estimation+correlation} to minimise the estimation error. In~\cite{MSE+OU+noisy}, the authors investigated the transmission of quantised and coded samples in a noisy OU model and showed that the minimum MSE can be presented by an increasing function of AoI. The concept of age of incorrect information (AoII) was proposed in~\cite{AoIIFirstPropose}, which incorporates the AoI and the error-based metric to measure the freshness of informative data. AoII-optimal sampling problems have been investigated for a binary Markov source~\cite{AoIIremoteestimation} and a semantics-based communication system~\cite{AoIISemantics}. Moreover, conditional entropy was used in~\cite{MIConditionalEntropy} to evaluate the staleness of data for estimation. In~\cite{SPAWC}, the mutual information was utilised to characterise the timeliness of information, and the authors studied the optimal sampling policy for Markov models. Despite these contributions, hidden Markov models have not been explicitly treated in related works.

In practical applications, the noise, interference, errors and other features can lead to severe performance degradation. This means that status updates generated at the source can be negatively affected, and may be hidden from observation when they are delivered to the receiver. However, existing works only treat Markov models in which variables are assumed to be directly visible at the destination node, and the timeliness of the system only relates to the most recent received status update.

Against this background, we are motivated to develop a general value of information (VoI) framework for hidden Markov models to characterise how valuable the status updates are at the receiver. In our previous work~\cite{mineGlobecom}, we defined the basic notion of the information value and commenced an investigation of VoI in the context of the Ornstein-Uhlenbeck (OU) process. The OU process is an important continuous, stationary and Gauss-Markov process which is able to represent various practical applications~\cite{introOU}. For example, it can be used to model the mobility of a drone that hovers at a fixed point but experiences positional disturbances in UAV networks. In this paper, we extend the basic model and go into more depth with regard to different sampling policies. The contributions of this paper are given as follows:
\begin{itemize}
    \item A VoI framework is formalised for hidden Markov models. The VoI is defined as the mutual information between the current status and a dynamic sequence of past observations, which gives a theoretical interpretation of the reduction in uncertainty in the current (unobserved) status of a hidden process given that we have noisy measurements.
    \item The VoI is explored in the context of one of the most important hidden Markov models: the noisy Ornstein-Uhlenbeck process, and its closed-form expressions are derived.
    \item The VoI with different sampling policies is investigated. Simplified VoI expressions are derived in both large and small noise regimes for arbitrary and uniform sampling times. The probability density and cumulative distribution of the worst-case VoI are analysed in a particular case: the M/M/1 queueing model.
    \item The optimal sampling strategy is studied to maximise the average VoI under the sampling rate constraint.
    \item Numerical results are provided to verify the theoretical analysis. The effect of noise, number of observations, sampling rate and correlation on VoI and its statistical properties are discussed. The performance of VoI for Markov and hidden Markov models are also presented.  
\end{itemize}

The remainder of this paper is organised as follows. The VoI formulation for hidden Markov models is given in Section II. The VoI for a specific hidden Markov model (the noisy OU process) is analysed in Section III. Effective sampling policies on VoI are explored in Section IV. The VoI-optimal sampling policy is studied in Section V. Numerical results and analysis are provided in Section VI. Conclusions are drawn in Section VII.
\section{Value of Information Formulation}
\subsection{Definition}
We consider a status update system where the source node continuously monitors a random process and samples data to get timely status updates of the targeted process, and these time-stamped messages will be transmitted via the communication system to the destination node for further analysis. As communication resources are limited, we assume that the transmission delay exists when status updates are received by the destination.

We denote $\{X_t\}$ as the random process under observation at the source node. Here, the time variable $t$ can be either continuous or discrete. Denote $(t_i,X_{t_i})$ as the message which is generated at arbitrary time $t_i$, and contains the corresponding value $X_{t_i}$ of the underlying random process. The status update is received by the destination node at time $t'_i$ with $t'_i > t_i$. The observations at the receiver are recorded in the observed random process $\{Y_t\}$ where $Y_{t'_i}$ is the observation corresponding to $X_{t_i}$. For the given time period $(0,t)$, denote $n$ as the index of the most recent data received at time $t'_n$ with $t'_n<t \le t'_{n+1}$.

In this paper, we define the value of information as the mutual information between the current status of the underlying random process at the source and a dynamic sequence of past observations captured by the receiver. For the given time instants, the general definition of VoI is given as 
\begin{equation}
\label{eq:general definition}
    v(t) = I({X_{t}};{Y_{t'_{n}}}, \cdots ,{Y_{t'_{n-m+1}}}), \quad t>t'_n,
\end{equation}
which is conditioned on times $\{t'_i\}$. Here, $n$ is the total number of recorded observations during the time period $(0,t)$. We look back in time and use a dynamic time window containing the most recent $m$ of $n$ samples ($1 \le m \le n$) to measure the information value of the current status $X_t$ of a hidden process. 
\subsection{VoI for Hidden Markov Models}
\begin{figure}
\centering
\includegraphics[width=8.5cm]{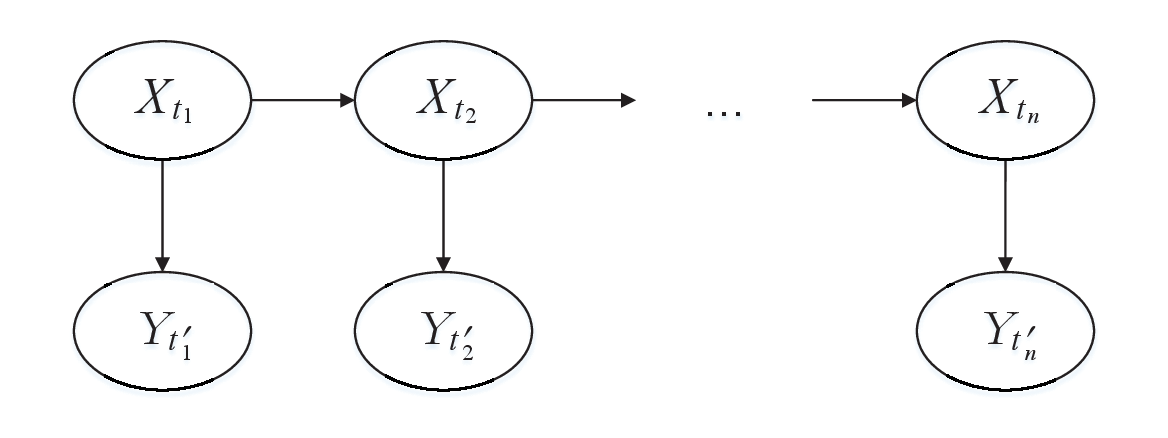}
\caption{Temporal evolution of hidden Markov models.} 
\label{fig:HMM}
\end{figure}

In the Markov model, the random process $\{X_t\}$ is directly visible, and the observations are also Markovian, i.e., $Y_{t_i'}=X_{t_i}$, for all $1 \le i \le n$. In this case, the VoI can be simplified to~\cite{SPAWC}
\begin{equation}
    v(t) = I(X_t;X_{t_n}),\quad t>t'_n.
\end{equation}
The VoI in the Markov model is independent of the number of observations $m$ and only depends on the most recent single status update. For hidden Markov models (Fig.~\ref{fig:HMM}), the observations at the receiver may be different from the initial value, i.e., $Y_{t_i'} \not= X_{t_i}$, but where
\begin{equation}
     \pr[Y_{t_i'}\in A | X_{t_1},\ldots,X_{t_i}] = \pr[Y_{t_i'}\in A | X_{t_i}]
\end{equation}
for all admissible $A$. Hence, the initial samples $\{X_{t_i}\}$ are invisible at the receiver. In this case, we have
\begin{equation}
    I({X_{t}};{Y_{t'_{n}}}, \cdots ,{Y_{t'_{n-m+1}}}) \ge I({X_{t}};{Y_{t'_{n}}}, \cdots ,{Y_{t'_{n-m+2}}}),
\end{equation}
and
\begin{equation}
\label{eq:MM upper bound}
\begin{aligned}
  v(t) &= h(X_t) - h(X_t|{Y_{t'_{n}}}, \cdots ,{Y_{t'_{n-m+1}}})\\
 &\le h(X_t) - h(X_t|{Y_{t'_{n}}}, \cdots ,{Y_{t'_{n-m+1}}},X_{t_n})\\
 &= h(X_t) - h(X_t|X_{t_n})\\
 &= I(X_t;X_{t_n})
\end{aligned}
\end{equation}
for $2 \le m \le n$. We find that the VoI increases with the number of observations $m$ and converges when more past observations are used. Moreover, the VoI in the Markov model can be regarded as the upper bound of the VoI in the hidden Markov model which illustrates that the lack of a direct route to observe $\{X_t\}$ reduces the information value.

The difference between the VoI in the Markov model and its counterpart in the hidden Markov model can be expressed as
\begin{equation}
\label{eq:correction}
\begin{aligned}
        & {I}({X_t};{X_{{t_n}}}) - {I}({X_t};{Y_{t{'_n}}}, \cdots ,{Y_{t{'_{n - m + 1}}}})\\
        = &h({X_t}|{Y_{t{'_n}}}, \cdots ,{Y_{t{'_{n - m + 1}}}}) - h({X_t}|{X_{{t_n}}})\\
        =& h({X_t}|{Y_{t{'_n}}}, \cdots ,{Y_{t{'_{n - m + 1}}}}) - h({X_t}|{X_{{t_n}}},{Y_{t{'_n}}}, \cdots ,{Y_{t{'_{n - m + 1}}}})\\
        = & I({X_t};{X_{{t_n}}}|{Y_{t{'_n}}}, \cdots ,{Y_{t{'_{n - m + 1}}}}).
\end{aligned}
\end{equation}
This reduction can be interpreted as the ``correction'' which captures the VoI gap due to the indirect observation in the hidden Markov model. In other words, we can think the VoI for the hidden Markov model as the VoI for the Markov model minus the correction. The ``correction'' can be quantified by the mutual information between the current status $X_t$ and the most recent (unobserved) status update $X_{t_n}$ conditioned on the knowledge of a sequence of past observations $\{{Y_{t{'_n}}}, \cdots ,{Y_{t{'_{n - m + 1}}}}\}$.

\section{VoI for a Noisy OU Process}
\subsection{Noisy OU Process Model}
In this section, we consider a particular case of a noisy Ornstein–Uhlenbeck process to show how the proposed VoI framework can be applied in the hidden Markov model. The underlying OU process $\{X_t\}$ satisfies the following stochastic differential equation (SDE)
\begin{equation}
\label{eq:OU SDE}
    \operatorname{d}\!X_t = \kappa (\theta-X_t) \operatorname{d}\!t + \sigma\operatorname{d}\!W_t
\end{equation}
where $\{W_t\}$ is standard Brownian motion, $\kappa$ is the rate of mean reversion, $\theta$ is the long-term mean, and $\sigma$ is the volatility of the random fluctuation. We assume that the initial value $X_0$ is normally distributed, specifically, $\mathcal{N}({\theta},\frac{\sigma^2}{2\kappa})$. 

For any $t$, the variable $X_t$ is also normally distributed with mean and variance:
\begin{equation}
    \E[{X_t}] = \theta, \quad \Var[X_t] = \frac{\sigma^2}{2\kappa}.
\end{equation}
Furthermore, $X_t$ conditioned on $X_s$ is Gaussian with mean and variance:
\begin{equation}
\begin{aligned}
    &\E[{X_t}|{X_s}] = \theta  + ({X_s} - \theta ){e^{ - \kappa (t - s)}},\\
    &\Var[{X_t}|{X_s}] = \frac{\sigma^2}{2\kappa}\left(1-e^{-2\kappa (t-s)}\right).
\end{aligned}
\end{equation}
The covariance of two variables is given by
\begin{equation}\label{eq:cov xts}
    \Cov[{X_t},{X_s}] = \frac{{{\sigma ^2}}}{{2\kappa }}{e^{ - \kappa |t - s|}}.
\end{equation}

We assume that the underlying OU process $\{X_t\}$ is observed through an additive noise channel. Therefore, this noisy OU model constitutes a hidden Markov model with observations defined as
\begin{equation}
     Y_{t_i'}=X_{t_i}+N_{t'_i}. 
\end{equation}
Here, $\{N_{t}\}$ is a noise process which is anchored at time $t'_i$ with the value $N_{t'_i}$. In practice, it can be used to represent the measurement or error that corrupts the status update $X_{t_i}$. We assume that $\{N_{t'_{i}}\}$ are independent and identically distributed (i.i.d.) Gaussian variables with zero mean and constant variance ${{\sigma}_n^2}$. Let the $m$-dimensional vector $\bm{X} = {[{X_{t_{n-m+1}}},\cdots,{X_{t_n}}]^{\operatorname{T}}}$ denote the sequence of status updates sampled by the source node, and its covariance matrix is given by
\begin{multline}
    \label{eq:cov x}
    {\mathbf{\Sigma_X} }= \\
    {\left[ {\begin{array}{*{20}{c}} {\Cov[{X_{{t_{n-m+1}}}},{X_{{t_{n-m+1}}}}]}& \cdots &{\Cov[{X_{{t_{n-m+1}}}},{X_{{t_n}}}]}\\
    \vdots & \ddots & \vdots \\
    {\Cov[{X_{{t_{n-m+1}}}},{X_{{t_{n}}}}]}& \cdots
    &{\Cov[{X_{{t_n}}},{X_{{t_n}}}]}
    \end{array}} \right]}.
\end{multline}
Let vector $\bm{Y} = {[Y_{t'_{n-m+1}},\cdots,Y_{t'_n}]^{\operatorname{T}}}$ denote the corresponding set of observations recorded at the receiver. Similarly, the associated noise samples are captured in vector $\bm{N} = {[N_{t'_{n-m+1}},\cdots,{N_{t'_n}}]^{\operatorname{T}}}$ with the covariance matrix 
\begin{equation}
    \label{eq:cov n}
        \mathbf{\Sigma_N}=\sigma_n^2\mathbf{I},
\end{equation}
where $\mathbf{I}$ is the identity matrix. Therefore, the observations of the noisy OU process can be collectively represented by
\begin{equation}
\label{eq:OU hidden mapping}
     \bm{Y}=\bm{X}+\bm{N}.
\end{equation} 
\subsection{VoI for the Noisy OU Process}
Based on the model given before, we can state the following main result of this section.
\begin{proposition}
\label{prop:1}
    Let the $m$-dimensional matrix $\mathbf{A} = \sigma_n^2\mathbf{\Sigma}^{ - 1}_{\mathbf{X}} +\mathbf{I}$, and denote $\mathbf A_{ij}$ as the $(m-1)\times (m-1)$ matrix constructed by removing the $i$th row and the $j$th column of the matrix $\mathbf A$. The VoI for the noisy OU process defined above can be written as
    \begin{multline}
    \label{eq:prop. 1 general voi}
          v(t) =  \frac{1}{2}\log \bigg(\frac{{1 }}{{1 - {e^{ - 2\kappa (t - {t_n})}}}}\bigg) \\
          - \frac{1}{2}\log \bigg(1 + \frac{{{1 }}}{{ \left(e^{2\kappa (t-t_n)} - 1\right) }} \frac{\det(\mathbf{A}_{mm})}{\gamma\det(\mathbf{A})}\bigg).
    \end{multline}
    Here, $\gamma$ is denoted as the ratio of the variance of the OU process and the variance of the noise, i.e.,
    \begin{equation}
    \label{eq:gamma snr}
    \gamma = \frac{\Var[X_{t_i}]}{\Var[N_{t'_i}]}=\frac{{\sigma ^2}}{{{2\kappa \sigma _n^2}}}.
    \end{equation}
\end{proposition}
\begin{IEEEproof}
    See appendix~\ref{appendix:voi} .
    \end{IEEEproof}
It is easy to show that the first logarithmic term in~\eqref{eq:prop. 1 general voi} represents the VoI for the Markov OU model $X_t$. The remainder quantifies a ``correction'' to the VoI of the hidden process that arises due to the indirect observation of the process through the noisy channel, and it evaluates the result of~\eqref{eq:correction} in the example of the OU model. Note that both $\mathbf{A}$ and $\mathbf{A}_{mm}$ are positive semi-definite, thus the second logarithmic term in~\eqref{eq:prop. 1 general voi} is non-negative. The parameter $\gamma$ gives a comparison between the randomness in the underlying OU process and the noise process in the communication channel, and it can be compared to the concept of the signal-to-noise ratio (SNR) in communication systems.
\subsection{Results for a Single Observation}

The result given in Proposition~\ref{prop:1} is general. In this subsection, we consider a special case ($m=1$) which gives the information about how much value the most recently received observation contains about the current status of a random process. In this case, the VoI can be calculated by replacing the $m$-dimensional vector $\bm{Y}$ with the single variable $Y_{t'_n}$ in~\eqref{eq:general definition}, which leads to the following corollary.
\begin{corollary}\label{cor:1}
The VoI for the noisy OU process with a single observation is given by
\begin{equation}
    \label{eq:general voi 1 dimension}
    v(t) 
    = - \frac{1}{2}\log \bigg(1 - \frac{\gamma }{{1 + \gamma }}{e^{ - 2\kappa (t - {t_n})}}\bigg).
\end{equation}
\end{corollary}
\begin{IEEEproof}
    This result follows directly from Proposition~\ref{prop:1} where $\det(\mathbf{A}_{mm})\coloneqq 1$.
\end{IEEEproof}
This corollary shows that for fixed $t_n$, as time $t$ increases, the VoI will decrease and the newly received update can cause a corresponding reset of $v(t)$. This is somewhat similar to the concept of AoI, which is equal to $t'_n-t_n$ at the moment the $n$th update arrives and then increases with unit slope until the next update comes. However, the VoI will decrease like $O(e^{-2\kappa t})$ until a new status update is received. The parameter $\kappa$ can be used to represent how correlated the updates are. Therefore, compared with AoI, the proposed VoI framework not only reflects the time evolution of a random process, but also captures the correlation properties of the underlying data source and the noise in the transmission channel.
\section{Effective Sampling Policies for VoI}
Having derived the general VoI framework, Corollary~\ref{cor:1} looks at the special case when the number of observations $m=1$ to illustrate the VoI concept clearly. When $m>1$, the covariance matrix $\mathbf{\Sigma_X}$ given in Proposition~\ref{prop:1} is closely related to the sampling times of the status updates. Therefore, it is important to take a view of different sampling policies and study how the sampling time affects the VoI. In this section, we consider three cases. First, we start with a general but fixed sampling policy. Then, we move to a uniform sampling policy. We conclude with a Poisson sampling policy, which is effectively the M/M/1 model.
\subsection{Arbitrary Fixed Sampling Intervals}
We assume that status updates are generated at arbitrary but fixed times $\{t_i\}$, and denote $T_i$ as the sampling interval of two packets, i.e.,
\begin{equation}
\label{eq:T_i}
    {T_i} = {t_{n-m+i+1}} - {t_{n-m+i}}, \quad 1 \le i \le m-1.
\end{equation}
In this case, the covariance matrix of $\bm{X}$ can be written as
\begin{equation}
    \mathbf{\Sigma _X} = \frac{{{\sigma ^2}}}{{2\kappa }}\left[ {\begin{array}{*{20}{c}}
1&{{e^{ - \kappa {T_1}}}}& \cdots &{{e^{ - \kappa \sum\limits_{i = 1}^{m-1} {{T_i}} }}}\\
{{e^{ - \kappa {T_1}}}}&1& \cdots &{{e^{ - \kappa \sum\limits_{i = 2}^{m-1} {{T_i}} }}}\\
 \vdots & \vdots & \ddots & \vdots \\
{{e^{ - \kappa \sum\limits_{i = 1}^{m-1} {{T_i}} }}}&{{e^{ - \kappa \sum\limits_{i = 2}^{m-1} {{T_i}} }}}& \cdots &1
\end{array}} \right].
\end{equation}
For simplicity, let
\begin{equation}
    {R_i} = \frac{1}{{1 - {e^{ - 2\kappa {T_i}}}}}, \quad 1 \le i \le m-1.
\end{equation}
The inverse of the covariance matrix of $\bm{X}$ is tridiagonal, which can be written as~\cite{irregularAR1,TridiPoissonIndepen}
\begin{equation}
\label{eq: Sigma_x inverse general}
    \mathbf{\Sigma}^{ - 1}_{\mathbf{X}} = \frac{{2\kappa }}{{{\sigma ^2}}}\left[ {\begin{array}{*{20}{c}}
{{a_1}}&{{b_1}}&{}&{}&{}\\
{{b_1}}&{{a_2}}&{{b_2}}&{}&{}\\
{}&{{b_2}}& \ddots & \ddots &{}\\
{}&{}& \ddots &{{a_{m - 1}}}&{{b_{m - 1}}}\\
{}&{}&{}&{{b_{m - 1}}}&{{a_m}}
\end{array}} \right],
\end{equation}
where 
\begin{equation}
    {a_i} = \left\{ {\begin{array}{*{20}{l}}
{{R_1}}&{i = 1}\\
{{R_{m-1}}}&{i = m}\\
{{R_{i-1}} + {R_{i }} - 1}&{\text{others}}
\end{array}} \right.
\end{equation}
and
\begin{equation}
    b_i =-\sqrt{ {R_{i }}({R_{i }} - 1)}, \quad 1 \le i \le m-1.
\end{equation}
Then, the matrix $\mathbf{A}$ in Proposition~\ref{prop:1} can be written as
\begin{multline}
\label{eq: poi A}
     \mathbf{A} = \sigma _n^2\mathbf{\Sigma}^{ - 1}_{\mathbf{X}} + \mathbf{I}=\\
     \left[ {\begin{array}{*{20}{c}}
{{{\frac{1}{\gamma}}a_1+1}}&{{{\frac{1}{\gamma}}b_1}}&{}&{}&{}\\
{{{\frac{1}{\gamma}}b_1}}&{{{\frac{1}{\gamma}}a_2+1}}&{{{\frac{1}{\gamma}}b_2}}&{}&{}\\
{}&{{{\frac{1}{\gamma}}b_2}}& \ddots & \ddots &{}\\
{}&{}& \ddots &{{{\frac{1}{\gamma}}a_{m - 1}+1}}&{{{\frac{1}{\gamma}}b_{m - 1}}}\\
{}&{}&{}&{{{\frac{1}{\gamma}}b_{m - 1}}}&{{{\frac{1}{\gamma}}a_m+1}}
\end{array}} \right].
\end{multline}

Thus, we have a clear and general presentation of the matrix $\mathbf{A}$  for arbitrary sampling intervals. The SNR parameter $\gamma$ in this matrix can significantly affect the VoI in the hidden Markov model. If $\gamma$ is large, the underlying latent process is dominant; otherwise, the noise process is dominant. Therefore, it is interesting to explore the VoI expressions with different levels of noise.

First, we consider the high SNR regime in which the small variance of noise leads to large $\gamma$, i.e., $\frac{1}{\gamma} \to 0$. In this case, we can state the following result.
\begin{lemma}
\label{lemma: induction}
Let $f_i$ denote the determinant of the $i$-dimensional matrix constructed from the first $i$ columns and rows of matrix $\mathbf{A}$, i.e., $f_{m-1}=\det(\mathbf{A}_{mm})$ and $f_{m}=\det(\mathbf{A})$. In the high SNR regime, $f_k$ can be calculated as
\begin{equation}
    \label{eq:f_m}
    {f_k}  = 1+\sum\limits_{i = 1}^k {{a_i}{\frac{1}{\gamma}} + \bigg(\sum\limits_{1 \le i < j \le k} {{a_i}{a_j} - \sum\limits_{i = 1}^{k - 1} {b_i^2} } \bigg){{\frac{1}{\gamma^2}}} + O{{(\frac{1}{\gamma^3})}}}
\end{equation}
for $1 < k \le m$.
\end{lemma}
\begin{IEEEproof}
    See appendix~\ref{appendix:induction}.
\end{IEEEproof}
Now, we have the following corollary.
\begin{corollary}
\label{cor: random high snr}
For the noisy OU process with arbitrary but fixed sampling times, the VoI in the high SNR regime can be written as
\begin{multline}
    \label{eq:poi approx VoI high snr}
         v(t) = \frac{1}{2}\log \bigg(\frac{1}{{1 - {e^{ - 2\kappa (t - {t_n})}}}}\bigg) - \frac{1}{2}\log \bigg[1 + \frac{1}{{{e^{2\kappa (t - {t_n})}} - 1}}\\
         \times \bigg( \frac{1}{\gamma } - \frac{1}{{1 - {e^{ - 2\kappa ({t_n} - {t_{n - 1}})}}}}\frac{1}{{{\gamma ^2}}}  \bigg)\bigg]+ O(\frac{1}{{{\gamma ^3}}}).
\end{multline}
\end{corollary}
\begin{IEEEproof}
For simplicity, we temporarily denote the coefficient of the second-order term $\frac{1}{\gamma^2}$ in~\eqref{eq:f_m} as $c_k$, i.e.,
\begin{equation}
    c_k=\sum\limits_{1 \le i < j \le k} {{a_i}{a_j} - \sum\limits_{i = 1}^{k - 1} {b_i^2} }.
\end{equation}
Based on Lemma~\ref{lemma: induction}, we have
\begin{equation}
\begin{aligned}
\label{eq:poi high snr ratio}
     \frac{\det(\mathbf{A}_{mm})}{\gamma\det(\mathbf{A})}&=\frac{1}{\gamma }\cdot\frac{{1 + \sum\limits_{i = 1}^{m - 1} {{a_i}\frac{1}{\gamma }}  + {c_{m - 1}}\frac{1}{{{\gamma ^2}}} + O(\frac{1}{{{\gamma ^3}}})}}{{1 + \sum\limits_{i = 1}^m {{a_i}\frac{1}{\gamma }}  +{c_m}\frac{1}{{{\gamma ^2}}} + O(\frac{1}{{{\gamma ^3}}})}}\\
     &= \frac{1}{\gamma } - {a_m}\frac{1}{{{\gamma ^2}}} + (b_{m - 1}^2 - a_m^2)\frac{1}{{{\gamma ^3}}} + O(\frac{1}{{{\gamma ^4}}})\\
      &= \frac{1}{\gamma } - {R_m}\frac{1}{{{\gamma ^2}}} - {R_m}\frac{1}{{{\gamma ^3}}} + O(\frac{1}{{{\gamma ^4}}})\\
      &= \frac{1}{\gamma } - \frac{1}{1-e^{-2\kappa(t_n-t_{n-1})}}\frac{1}{{{\gamma ^2}}} + O(\frac{1}{{{\gamma ^3}}}).
     \end{aligned}
\end{equation}
The result of this corollary given in~\eqref{eq:poi approx VoI high snr} is obtained by substituting~\eqref{eq:poi high snr ratio} into~\eqref{eq:prop. 1 general voi}.
\end{IEEEproof}

Similar to Proposition~\ref{prop:1}, the first logarithmic term in Corollary~\ref{cor: random high snr} represents the VoI for the non-noisy Markov OU process $\{X_t\}$, and the remainder quantifies the ``correction''. The expression of $v(t)$ is conditioned on the past time instants $t_n$ and $t_{n-1}$, and the VoI does not depend on $m$ (the number of samples). This is because when $\gamma$ is large, the Markov OU randomness is dominant, and the noisy channel is not expected to play a vital role in the calculation of VoI. Therefore, the VoI in the high SNR regime approaches its Markov counterpart, which is not related to $m$. 

Moreover, if the VoI given in~\eqref{eq:poi approx VoI high snr} is truncated to the second-order term $\frac{1}{\gamma^2}$, the approximated VoI will first decrease and then increase with $\frac{1}{\gamma}$. The turning point is at $\frac{1}{\gamma}=\frac{1}{2}(1-e^{-2\kappa(t_n-t_{n-1})})$. Therefore, the valid region of the approximated VoI is
\begin{equation}
\label{eq:valid region poi high}
    \gamma \ge \frac{2}{1-e^{-2\kappa(t_n-t_{n-1})}}.
\end{equation}

Next, we study the VoI expression in the low SNR regime (i.e., $\gamma \to 0$). We can state the following result.
\begin{corollary}
\label{cor: poi low snr}
For the noisy OU process with arbitrary but fixed sampling times, the VoI in the low SNR regime can be written as
\begin{multline}
    \label{eq:poi appro. low snr}
    v(t) =  - \frac{1}{2}\log \bigg[1 - {e^{ - 2\kappa (t - {t_n})}}\bigg(1 + \sum\limits_{j = 1}^{m - 1} {{e^{ - 2\kappa \sum\limits_{i = j}^{m - 1} {{T_i}} }}} \bigg)\gamma \bigg] \\+ O({\gamma ^2})
\end{multline}
\end{corollary}
\begin{IEEEproof}
  See appendix~\ref{appendix:cor general low snr}.
\end{IEEEproof}

The VoI in the low SNR regime is conditioned on all past time instants. The randomness of the noise dominates in the low SNR regime, thus the VoI relates to the number of observations $m$ and increases as $m$ grows larger. Compared with Corollary~\ref{cor: random high snr}, the VoI in the low SNR regime can be improved by using more past observations, while the number of observations $m$ is not able to help to improve the VoI in the high SNR regime.

\subsection{Uniform Sampling Intervals}
Corollaries~\ref{cor: random high snr} and~\ref{cor: poi low snr} give general results for the VoI with arbitrary sampling times in high and low SNR regimes, respectively. In some cases, status updates are generated at a fixed rate. It may still be of interest to have a clear representation of VoI with uniform sampling intervals. In this subsection, we consider this specific case in which status updates are sampled at regular times.

Let the sampling time $t_i=i\Delta t$ where the constant $\Delta t$ ($\Delta t>0$) denotes the fixed sampling interval for all $1 \le i \le n$. For simplicity, we use $\rho$ to denote $e^{ - \kappa \Delta t}$In this case, we are able to derive the closed-form expression of the determinant ratio which is given in~\eqref{eq:uni B determiant ratio}. The proof of this result is given in appendix~\ref{appendix:characteristic equation}. This means that we can obtain the general closed-form expression of the VoI given in Proposition~\ref{prop:1} when sampling times are regular. In addition to the closed-form expression, we are also able to obtain approximated VoI expressions in the high and low SNR regimes by substituting $T_i=\Delta t$ in Corollaries~\ref{cor: random high snr} and~\ref{cor: poi low snr}. We have the following corollaries.
\begin{figure*}[ht]
   \begin{equation}
   \begin{aligned}
   \label{eq:uni B determiant ratio}
    \frac{{\det ({\mathbf{A}_{mm}})}}{{\gamma\det (\mathbf{A})}} &=   \frac{1-\rho^2}{\rho}\cdot\frac{{ac(\lambda _1^{m - 2} - \lambda _2^{m - 2}) + (ab + bc)(\lambda _1^{m - 3} - \lambda _2^{m - 3}) + {b^2}(\lambda _1^{m - 4} - \lambda _2^{m - 4})}}{{{a^2}(\lambda _1^{m - 1} - \lambda _2^{m - 1}) + 2ab(\lambda _1^{m - 2} - \lambda _2^{m - 2}) + {b^2}(\lambda _1^{m - 3} - \lambda _2^{m - 3})}}\\
     \end{aligned}
    \end{equation}
\end{figure*}

\begin{corollary}
\label{cor: uni high snr}
For the noisy OU process with uniform sampling intervals, the VoI in the high SNR regime can be written as
\begin{multline}
    \label{eq:uni appro. high snr}
        v(t)= \frac{1}{2}\log \bigg(\frac{1}{{1 - {e^{ - 2\kappa (t - {t_n})}}}}\bigg) \\
        - \frac{1}{2}\log \bigg[1 + \frac{1}{{{e^{2\kappa (t - {t_n})}} - 1}}      \bigg(\frac{1}{\gamma } - \frac{1}{{(1 - {\rho ^2}){\gamma ^2}}}\bigg)\bigg]+ O(\frac{1}{{{\gamma^3 }}}).
\end{multline}
\end{corollary}
The valid region of the approximated VoI is
\begin{equation}
\label{eq:valid region uni high}
    \gamma \ge \frac{2}{1-\rho^2}.
\end{equation}

\begin{corollary}
\label{cor: uni low snr}
For the noisy OU process with uniform sampling intervals, the VoI in the low SNR regime can be written as
\begin{equation}
    \label{eq:uni appro. low snr}
    \begin{aligned}
     v(t) &=  - \frac{1}{2}\log \bigg[1 - \frac{{{{e ^{{-2\kappa(t - t_n)}}}}(1 - {\rho ^{2m}})}}{{1 - {\rho ^2}}}\gamma\bigg]+ O({\gamma ^2}).
    \end{aligned}
\end{equation}
\end{corollary}
For uniform sampling intervals, these corollaries further illustrate that the number of observations has a very small effect on the VoI in the high SNR regime. While in the low SNR regime, $m$ has a larger effect. As $0 < \rho <1$, the VoI converges to $- \frac{1}{2}\log [1 - \frac{{{{e ^{{-2\kappa(t - t_n)}}}}}}{{1 - {\rho ^2}}}\gamma]$ when $m$ grows to infinity in the low SNR regime. 

\subsection{Application of VoI in an M/M/1 System}
In this subsection, we consider the case of a first-come-first-serve (FCFS) M/M/1 queueing system (i.e., sampling times are Poisson distributed) to explore the statistical properties of VoI.  This investigation provides insight into potential applications of the proposed framework. 

In the M/M/1 model, status updates are assumed to be sampled as a rate $\lambda$ Poisson process and the service rate is $\mu$. The sampling interval of two packets
\begin{equation}
    {T_i} = {t_{i}} - {t_{i-1}}, \quad n-m+2 \le i \le n
\end{equation}
is an i.i.d. exponentially distributed random variable with mean $\frac{1}{\lambda }$ and variance $\frac{1}{{{\lambda ^2}}}$. Let the random variables $\{{W_i}\}$ ($ n-m+1 \le i \le n$) denote the service times, which are i.i.d. exponential random variables with mean $\frac{1}{\mu}$ and variance $\frac{1}{{{\mu ^2}}}$. Let the random variable
\begin{equation}
    {S_i} = {t'_{i}} - {t_{i}}, \quad n-m+1 \le i \le n
\end{equation}
represent the system time of the $i$th status update. When the system reaches steady state, the system times are also i.i.d. exponential random variables with mean $1/(\mu-\lambda)$~\cite{2012infocom?,bookRandomProcessSystemTime}.

We consider the case when $m=1$ and the VoI expression is given in Corollary~\ref{cor:1}. We observe that the VoI immediately reaches the local minimum before a new update is received by the destination. Given that $n$ samples are observed, the worst-case VoI can be obtained when $t=t'_{n+1}$.  This operating point is of interest in applications with a threshold restriction on the information value. When the time instants are random, the VoI in the worst case can also be regarded as a random variable. Based on~\eqref{eq:general voi 1 dimension}, the worst-case VoI with $n$ status updates is given by
\begin{equation}
\begin{aligned}
    V_n &= - \frac{1}{2}\log \bigg(1 - \frac{\gamma }{{1 + \gamma }}{e^{ - 2\kappa (t'_{n+1} - {t_n})}}\bigg)\\
     &=  - \frac{1}{2}\log \bigg(1 - \frac{\gamma }{{1 + \gamma }}{e^{ - 2\kappa ((t{'_{n + 1}} - {t_{n + 1}}) + ({t_{n + 1}} - {t_n}))}}\bigg)\\
     & =  - \frac{1}{2}\log \bigg(1 - \frac{\gamma }{{1 + \gamma }}{e^{ - 2\kappa ({S_{n + 1}} + {T_{n + 1}})}}\bigg).
\end{aligned}
\end{equation}

The system time $S_{n + 1}$ and the sampling interval $T_{n+1}$ are the main factors affecting the distribution of VoI. As shown in appendix~\ref{appendix:mm1 joint PDF}, the joint probability density function (PDF) of $S_{n + 1}$ and $T_{n+1}$ is given by
\begin{equation}
    {f_{T,S}}(t,s) = \lambda \mu {e^{ - \lambda t - \mu s}} - {\mu ^2}{e^{ - \mu (t + s)}} + \mu (\mu  - \lambda ){e^{ - \mu t - (\mu  - \lambda )s}}.
\end{equation}

Let the variable $Z_{n+1}=S_{n + 1}+T_{n+1}$ such that its PDF is given by
\begin{multline}
    \label{eq:PDF z}
    {f_Z}(z) =\int_0^z {{f_{T,S}}(z - s,s)} \operatorname{d}\!s =\mu \bigg[\frac{\lambda }{{\mu  - \lambda }}{e^{ - \lambda z}} -\\ \bigg(\frac{\lambda }{{\mu  - \lambda }} + \mu z + \frac{{\mu  - \lambda }}{\lambda }\bigg){e^{ - \mu z}} + \frac{{\mu  - \lambda }}{\lambda }{e^{ - (\mu  - \lambda )z}}\bigg].
\end{multline}
Define the monotonic function
\begin{equation}
    g(z) =  - \frac{1}{2}\log (1 - \frac{\gamma }{{1 + \gamma }}{e^{ - 2\kappa z}}).
\end{equation}
Then, the PDF of $V_n$ can be calculated as
\begin{equation}
\label{eq:v PDF mapping}
    {f_V}(v) = {f_Z}({g^{ - 1}}(v))\left| {\frac{\operatorname{d}}{{\operatorname{d}\!v}}({g^{ - 1}}(v))} \right|.
\end{equation}
    Here, the $g^{ - 1}$ denotes the inverse function, and we have
    \begin{equation}
    \label{eq:inverse v}
        {g^{ - 1}}(v) =  - \frac{1}{{2\kappa }}\log \bigg(\frac{{(1 + \gamma )(1 - {e^{ - 2v}})}}{\gamma }\bigg),
    \end{equation}
    \begin{equation}
    \label{eq:dervi v}
        \frac{\operatorname{d}}{{\operatorname{d}\!v}}({g^{ - 1}}(v)) =  - \frac{{{e^{ - 2v}}}}{{\kappa (1 - {e^{ - 2v}})}}.
    \end{equation}
Now, we can state the following results.
\begin{proposition}\label{prop:2}
    In the FCFS M/M/1 queueing system, the probability density function and the cumulative distribution function of the worst-case VoI are given by
\begin{multline}
\label{eq:PDF prop2}
{f_V}(v) = \frac{{\mu {e^{ - 2v}}}}{{\kappa (1 - {e^{ - 2v}})}}\bigg[\frac{\lambda }{{\mu  - \lambda }}{(r(v))^{\frac{\lambda }{{2\kappa }}}}-\bigg(\frac{\lambda }{{\mu  - \lambda }}\\
+ \frac{{\mu  - \lambda }}{\lambda } - \frac{\mu }{{2\kappa }}\log (r(v))\bigg){(r(v))^{\frac{\mu }{{2\kappa }}}} + \frac{{\mu  - \lambda }}{\lambda }{(r(v))^{\frac{{\mu  - \lambda }}{{2\kappa }}}}\bigg],
\end{multline}
and
\begin{multline}
\label{eq:CDF prop3}
    {F_V}(v)    =\frac{{\mu   }}{\mu- \lambda } {r(v)^{\frac{\lambda}{2\kappa}}}+\frac{\mu}{\lambda}{r(v)^{\frac{\mu-\lambda}{2\kappa}}}\\
    + \bigg(1-\frac{\mu^2 }{\lambda(\mu-\lambda) }+\frac{\mu}{2\kappa}\log r(v)\bigg) {r(v)^{\frac{\mu}{2\kappa}}}.
\end{multline}
Here, $r(v)$ is
\begin{equation}
    r(v) = \frac{{(1 + \gamma )(1 - {e^{ - 2v}})}}{\gamma }.
\end{equation}
\end{proposition}
\begin{IEEEproof}
The density function is obtained directly by substituting~\eqref{eq:PDF z}, ~\eqref{eq:inverse v} and~\eqref{eq:dervi v} into~\eqref{eq:v PDF mapping}. The cumulative distribution function (CDF) is obtained by 
\begin{equation}
    {F_V}(v) = \pr(V \le v) =\int_0^v {{f_V}(x)\operatorname{d}\!x}.
\end{equation}
\end{IEEEproof}
In practice, this distribution function can be interpreted as the ``VoI outage'', i.e., the probability that the VoI right before a new sample arrives is below a threshold $v$, which can play a vital role in the system design.
\section{Optimal Sampling Policy}
In the AoI literature, the optimal sampling policy has been studied in
~\cite{OptimalPolicyCDC2018,OptimalPolicyALLERTON2019,OptimalPolicyISIT2020,QueueSYUpdatWait,estimationWienerProcess,estimationOUprocess}. This is done by formulating a constrained optimisation problem to minimise the average AoI under the sampling rate constraint. In this section, we formulate and analyse this problem in the context of the VoI case. Our objective is to maximise the average VoI by optimising the sampling times under the maximum sampling rate constraint. Denoting the maximum allowable sampling rate as $f_{\max}$, this optimisation problem is formulated as
\begin{equation}
\begin{aligned}
\sup_{\{t_i\}} \quad & {\mathop {\liminf }\limits_{n \to \infty } \frac{{\E\bigg[\int_0^{t{'_n}} {v (t)} \operatorname{d}\!t\bigg]}}{{\E[t{'_n}]}} }\\
\textrm{s.t.} \quad & \mathop {\liminf }\limits_{n \to \infty } \frac{1}{n}\E[{t'_n}] \ge \frac{1}{{{f_{\max}}}}.    \\
\end{aligned}
\end{equation}

We consider the single observation case, in which the VoI in Corollary~\ref{cor:1} can be presented as a function of AoI, i.e.,
\begin{equation}
\label{eq:v(a) single}
    V(a)  = - \frac{1}{2}\log \bigg(1 - \frac{\gamma }{{1 + \gamma }}{e^{ - 2\kappa a}}\bigg).
\end{equation}
In order to obtain the highest information value, the next status update should be sampled after the previous sample is received, i.e., $t_{i+1} \ge t'_i$. For the $i$th status update, denote $S_i$ as the transmission delay
\begin{equation}
    {S_i} = {t'_{i}} - {t_{i}},
\end{equation}
and $Z_i$ as the waiting time
\begin{equation}
    {Z_i} = {t_{i+1}} - {t'_{i}}
\end{equation}
where $Z_i \ge 0$. We assume that transmission delays $\{S_i\}$ are i.i.d. random variables and the waiting time $Z_i$ relates to $S_i$. The average VoI with a single observation can be written as
\begin{equation}
    \begin{aligned}
    \E[V] &= \mathop {\lim }\limits_{n \to \infty } \frac{{\E\bigg[\int_0^{t{'_n}} v(t) \operatorname{d}\!t\bigg]}}{{\E[t{'_n}]}}  \\
   &=\mathop {\lim }\limits_{n \to \infty } \frac{{\E\bigg[\sum\limits_{i = 0}^{n - 1} {\int_{{S_i}}^{{S_i} + {z(S_i)} + {S_{i + 1}}} {V(a)} \operatorname{d}\!a} \bigg]}}{{\E\bigg[\sum\limits_{i = 0}^{n - 1} {({S_{i+1}} + {z(S_i)})} \bigg]}}\\
    & = \frac{{\E[q(S,z(S),S')]}}{{\E[S + z(S)]}},
     \end{aligned}
\end{equation}
where $S'$ has the same distribution of $S$ and the function $q(\cdot)$ is denoted as
\begin{equation}
    q(S,z(S),S') = \int_S^{S + z(S) + S'} { \frac{1}{2}\log \bigg(1 - \frac{\gamma }{{1 + \gamma }}{e^{ - 2\kappa a}}\bigg)} \operatorname{d}\!a.
\end{equation}
Thus, this optimisation problem can be further formulated as
\begin{equation}
\label{eq: problem formulation}
\begin{aligned}
\min_{z} \quad & {\frac{{\E[q(S,z(S),S')]}}{{\E[S + z(S)]}}}\\
\textrm{s.t.} \quad & {\E[S + z(S)] \ge \frac{1}{{{f_{\max }}}}}.    \\
\end{aligned}
\end{equation}

We aim to find the optimal waiting time $z(S_i)$ given $S_i$ to maximise the average VoI under the sampling rate constraint. The optimal solution for this problem is a threshold-based policy which is given in the following theorem.
\begin{theorem}[Theorem 5.5,~\cite{bookAoI}]
\label{theo:sampling policy}
The optimal waiting time to maximise the average VoI under the sampling rate constraint is given by
\begin{multline}
\label{eq:voi-optimal waiting time}
     z(S_i) = \inf \bigg\{ z \ge 0:\\
     \E\bigg[ {- \frac{1}{2}\log \bigg(1 - \frac{\gamma }{{1 + \gamma }}{e^{ - 2\kappa (S_i+z(S_i)+S_{i+1})}}\bigg)\bigg]} \le \beta \bigg\}.
\end{multline}
If $\E[S_i+z(S_i)] \ge \frac{1}{f_{\max}}$, then $\beta$ is the root of 
\begin{equation}
    \beta=\frac{{\E\bigg[{\int_{{S_i}}^{{S_i} + {z(S_i)} + {S_{i + 1}}} {{- \frac{1}{2}\log (1 - \frac{\gamma }{{1 + \gamma }}{e^{ - 2\kappa a}})}} \operatorname{d}\!a} \bigg]}}{{\E\bigg[ {{S_{i}} + {z(S_i})} \bigg]}};
\end{equation}
otherwise, $\beta$ is determined by solving
\begin{equation}
    \E[S_i+z(S_i)]=\frac{1}{f_{\max}}.
\end{equation}
\end{theorem}
Here, the waiting time $z(S_i)$ is conditioned on the transmission time $S_i$. The optimal policy for VoI has the same form as the AoI-optimal policy, i.e., they are both threshold-based policies. $\beta$ is the threshold which can be solved by the bisection search algorithm. Numerical results are presented in the following section to show the performance of this sampling policy.

\section{Numerical Results}
In this section, numerical results are provided through Monte Carlo simulations. Results show the VoI performance in Markov and hidden Markov models, verify the validity of the simplified VoI in the high and low SNR regimes and illustrate the effects brought about by altering the number observations, the sampling rate, the correlation and the noise. In the numerical experiments, the volatility parameter $\sigma$ of the OU model is set to $1$. We consider the FCFS transmission scheme, and the service time of each status update is generated randomly according to a rate $\mu=1$ exponential distribution. For arbitrary sampling times, the sampling process is chosen to be a rate $\lambda$ Poisson process.
\begin{figure}
\centering
\includegraphics[width=8.5cm]{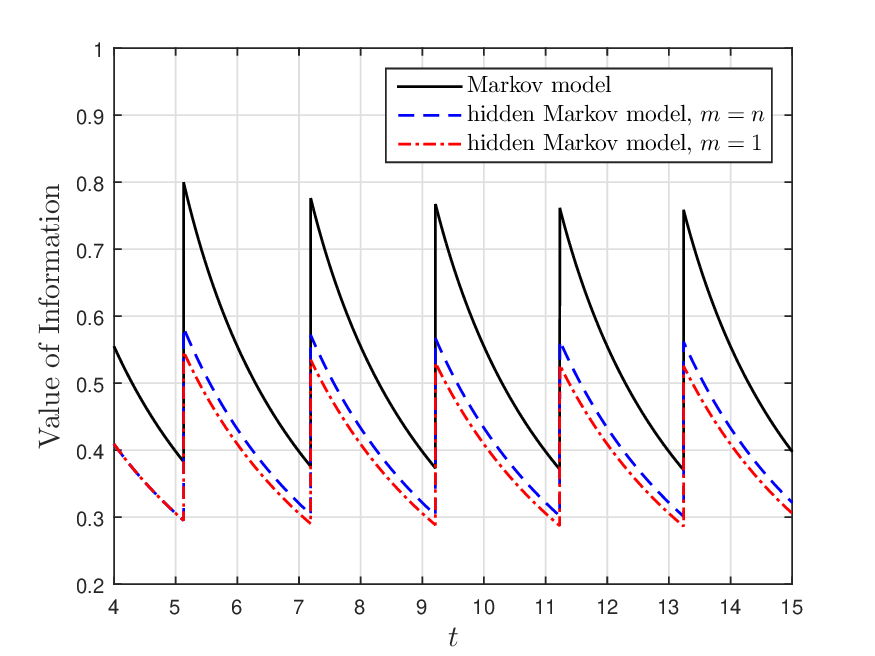}
\caption{Time evolution of VoI in Markov OU process and the noisy OU process; correlation parameter $\kappa=0.1$, noise parameter $\sigma_n^2=1$ and sampling interval $\Delta t=2$.}
\label{fig:voi time evolution}
\end{figure}

\begin{figure}
\centering
\includegraphics[width=8.5cm]{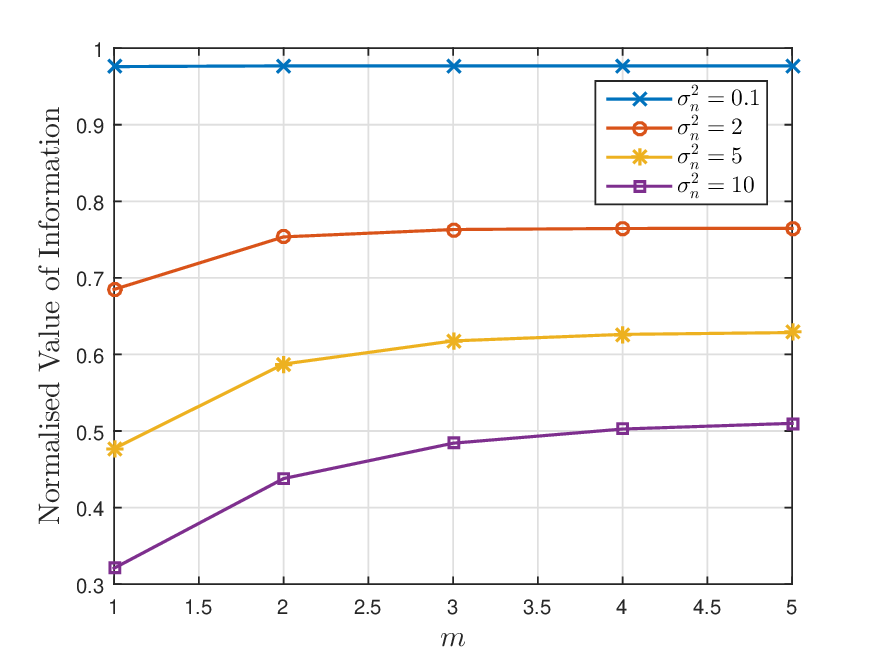}
\caption{VoI in the noisy OU process versus the number of observations $m$ for $\sigma_n^2 \in \{ 0.1,2,5,10\}$ at $t=100$; correlation parameter $\kappa=0.05$ and sampling interval $\Delta t=2$.}
\label{fig:voi-m}
\end{figure}
Fig.~\ref{fig:voi time evolution} shows the VoI in Markov and hidden Markov OU models for different numbers of observations $m$. In the figure, all the received observations are used for the result labelled ``$m=n$''; only the most recent received single observation is used for the result labelled ``$m=1$''. This figure verifies the results given in Proposition~\ref{prop:1} and Corollary~\ref{cor:1}. The VoI decreases with time until a new update is transmitted, which shows a behaviour that is similar to the traditional AoI evolution. The black curve represents the VoI in the underlying Markov OU model, which is the first term in Proposition~\ref{prop:1}. The gap between the result in the Markov model and its counterpart in the hidden Markov model is the second term in Proposition~\ref{prop:1}, which represents the ``VoI correction'' due to the indirect observation. The number of observations can affect the VoI in the hidden Markov model. However, the VoI in the Markov model does not depend on $m$.

Fig.~\ref{fig:voi-m} further shows how the VoI varies with the number of observations $m$ for different values of $\sigma_n^2$. The horizontal axis represents the number of observations we used to predict the value of the current status of the random process. The vertical axis is the normalised VoI, which represents the ratio of $v(t)$ to $v_\text{OU}(t)$, where $v_\text{OU}(t)$ is the VoI in the underlying Markov OU process. This result shows that the VoI in the noisy OU process increases with the number of observations, and converges as more past observations are used. This can be explained as more past observations can give more information about the current status of the latent random process. Moreover, the normalised VoI approaches $1$ for small $\sigma_n^2$, which means that the VoI in the Markov model can be regarded as the upper bound of its counterpart in the hidden Markov model~\eqref{eq:MM upper bound}.
\begin{figure}
\centering
\includegraphics[width=8.5cm]{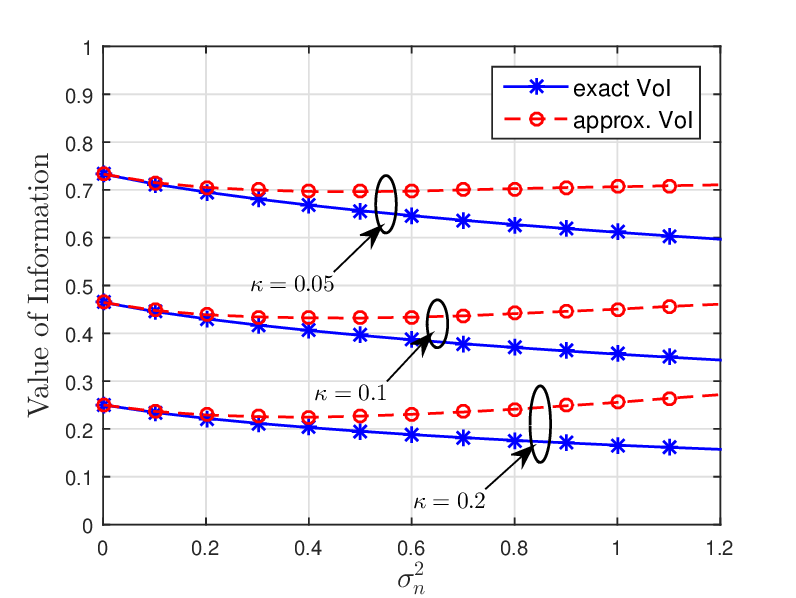}
\caption{High SNR regime: Comparison of the exact VoI and the approximated VoI with arbitrary sampling times for $\kappa \in \{ 0.05,0.1,0.2\}$ at $t=100$; sampling rate $\lambda=0.5$ and the number of observations $m=5$.}
\label{fig:poi high SNR}
\end{figure}

\begin{figure}
\centering
\includegraphics[width=8.5cm]{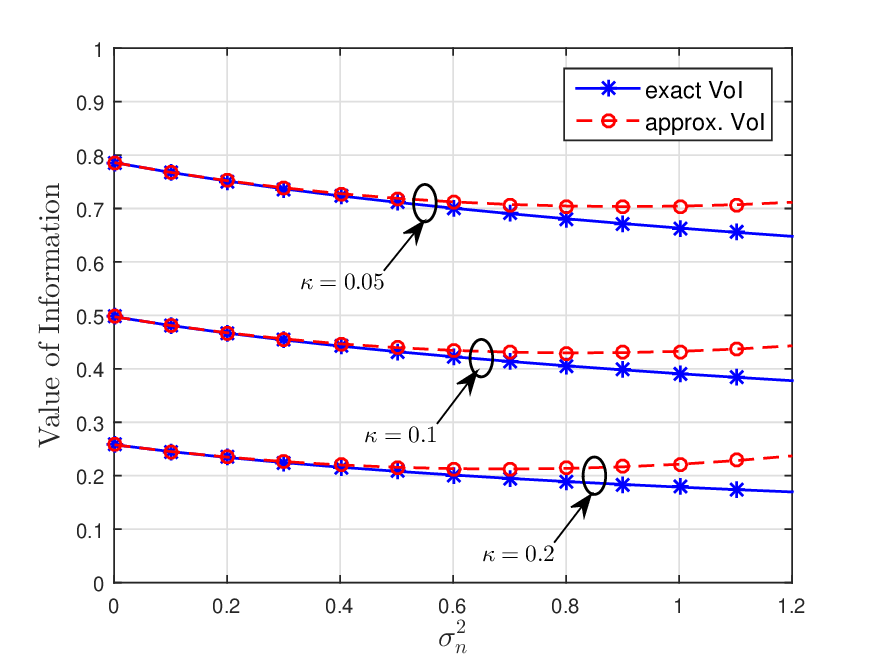}
\caption{High SNR regime: Comparison of the exact VoI and the approximated VoI with uniform sampling intervals for $\kappa \in \{ 0.05,0.1,0.2\}$ at $t=100$; sampling interval $\Delta t=2$ and the number of observations $m=5$.}
\label{fig:uni high SNR}
\end{figure}
\begin{figure}
\centering
\includegraphics[width=8.5cm]{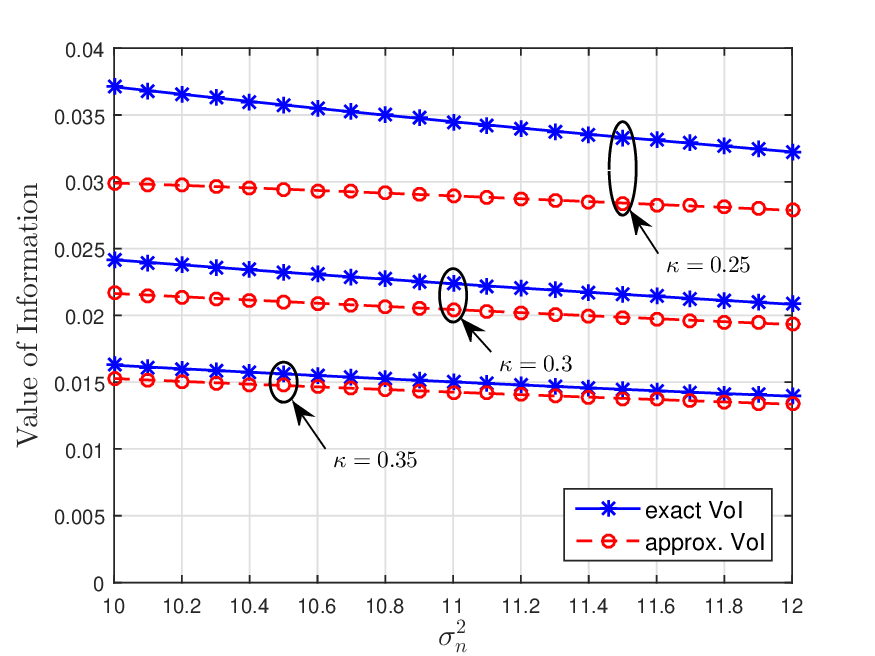}
\caption{Low SNR regime: Comparison of the exact VoI and the approximated VoI with uniform sampling intervals for $\kappa \in \{ 0.25,0.3,0.35\}$ at $t=100$; sampling interval $\Delta t=2$ and the number of observations $m=5$.}
\label{fig:uni low SNR}
\end{figure}

Figs.~\ref{fig:poi high SNR},~\ref{fig:uni high SNR} and~\ref{fig:uni low SNR} show the numerical validation of the exact general VoI given in Proposition~\ref{prop:1} and the approximated VoI with different sampling policies in different SNR regimes which are discussed in Corollaries~\ref{cor: random high snr},~\ref{cor: uni high snr} and~\ref{cor: uni low snr}. Figs.~\ref{fig:poi high SNR} and~\ref{fig:uni high SNR} consider the high SNR regime with arbitrary and uniform sampling intervals, respectively. We compare the exact VoI given in~\eqref{eq:prop. 1 general voi} with the approximated VoI in the high SNR regime given in~\eqref{eq:poi approx VoI high snr} and~\eqref{eq:uni appro. high snr}, respectively. It is not surprising that the exact VoI decreases as $\sigma_n^2$ increases. As the approximated VoI is truncated to the second-order term of $\frac{1}{\gamma}$, the VoI first decreases and starts to increase when it reaches the invalid region as $\sigma_n^2$ increases. The turning points are $\{0.5,0.4,0.3\}$ in Fig.~\ref{fig:poi high SNR} and $\sigma_n^2 \in \{0.9,0.8,0.7\}$ in Fig.~\ref{fig:uni high SNR}, verifying the results given in~\eqref{eq:valid region poi high} and~\eqref{eq:valid region uni high}. As expected, the approximated VoI is very close to the actual VoI when $\sigma_n^2$ and $\kappa$ are small, while the gap increases when the system experiences larger noise. Fig.~\ref{fig:uni low SNR} considers the low SNR regime, and compares the exact VoI with the approximated VoI given in~\eqref{eq:uni appro. low snr}. Compared to the high SNR regime, we observe the opposite behaviour, i.e., the approximated VoI is approaching the exact VoI when $\sigma_n^2$ and $\kappa$ are large. Therefore, these simulation results verify the analysis in Corollaries~\ref{cor: random high snr},~\ref{cor: uni high snr}, and~\ref{cor: uni low snr}.

\begin{figure}
\centering
\includegraphics[width=8.5cm]{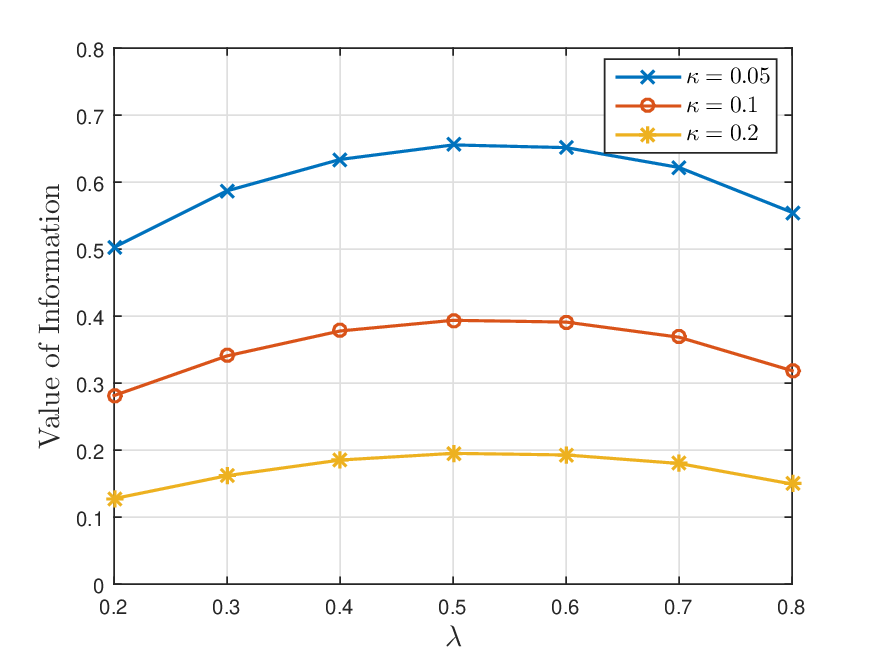}
\caption{VoI in the noisy OU process versus the sampling rate $\lambda$ for $\kappa \in \{ 0.05, 0.1, 0.2\}$ at $t=100$; noise parameter $\sigma_n^2=0.5$ and the number of observations $m=2$.}
\label{fig:voi-lambda}
\end{figure}

In Fig.~\ref{fig:voi-lambda}, we investigate the effect of the sampling rate and correlation on the VoI in the noisy OU process. Fixing $\kappa$, we observe that both small and large sampling rates lead to small VoI. For small $\lambda$, the system lacks the newly generated status updates to predict the current status of the underlying random process. For large $\lambda$, more status updates have been sampled at the source, but they may not be transmitted in a timely manner because they need to wait for a longer time in the FCFS queue before being transmitted. Fixing $\lambda$, the system sees the large value when $\kappa$ is small. The parameter $\kappa$ represents the mean reversion which can be used to capture the correlation of the latent OU process. Compared to the less correlated samples (larger $\kappa$), the value of highly correlated samples is larger, which further illustrates that ``old" samples from the highly correlated source may still have value in some cases.

\begin{figure}
\centering
\includegraphics[width=8.5cm]{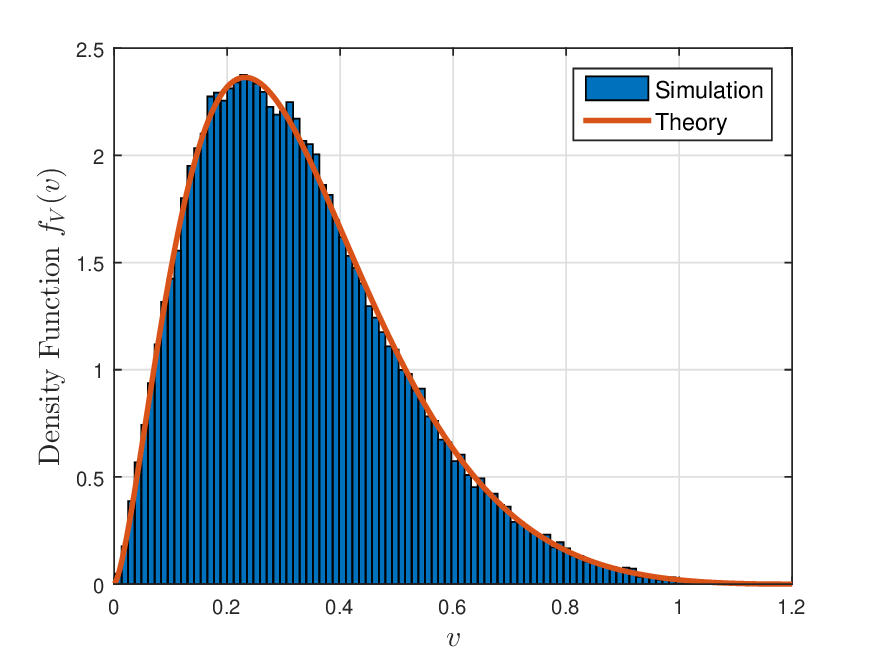}
\caption{The density function of the worst-case VoI; correlation parameter $\kappa=0.1$, noise parameter $\sigma_n^2=0.5$ and sampling rate $\lambda=0.5$.}
\label{fig:PDF}
\end{figure}

\begin{figure}
\centering
\includegraphics[width=8.5cm]{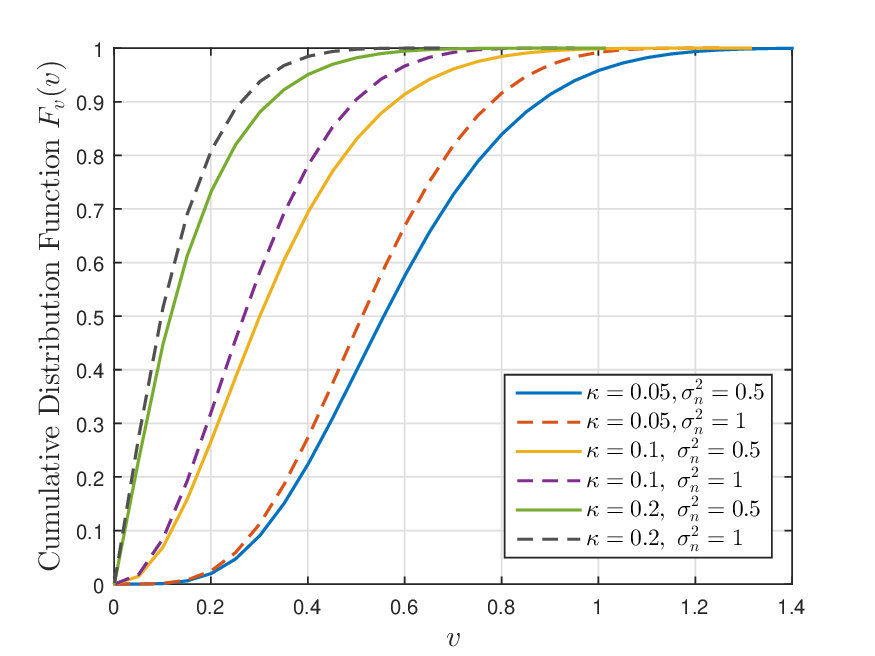}
\caption{The cumulative distribution function of the worst-case VoI for $\kappa \in \{ 0.05, 0.1, 0.2, 0.3\}$ and $\sigma_n^2 \in \{0.5,1\}$; sampling rate $\lambda=0.5$.}
\label{fig:CDF}
\end{figure}

Figs.~\ref{fig:PDF} and~\ref{fig:CDF} show statistical properties of the VoI in the worst case. Fig.~\ref{fig:PDF} shows the density of the discrete path of the worst-case VoI and the theoretical density function given in~\eqref{eq:PDF prop2} of Proposition~\ref{prop:2} when $\kappa=0.1$. It is clear to find that the results obtained from Monte Carlo simulations are consistent with the PDF obtained from the theoretical analysis. In Fig.~\ref{fig:CDF}, we plot the CDF of the worst-case VoI given in~\eqref{eq:CDF prop3} Proposition~\ref{prop:2} for different values of $\kappa$ and $\sigma_n^2$. This figure illustrates that the ``VoI outage" is more likely to occur when the status updates are less correlated or the system experiences large noise.

\begin{figure}
\centering
\includegraphics[width=8.5cm]{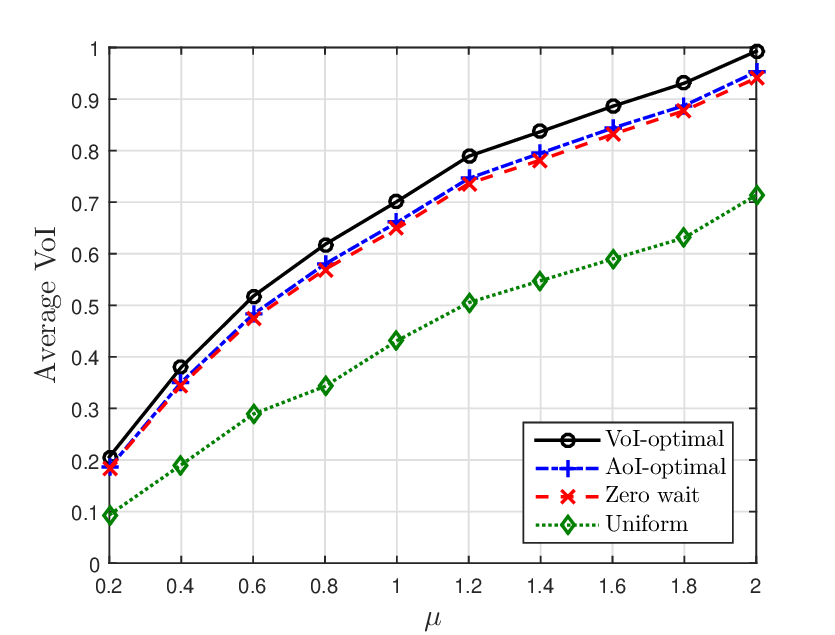}
\caption{Average VoI in the noisy OU process versus the transmission rate parameter $\mu$ for different sampling policies without sampling rate constraints; correlation parameter $\kappa=0.1$ and the noise parameter $\sigma_n^2=0.1$.}
\label{fig:sampling voi vs mu }
\end{figure}

\begin{figure}
\centering
\includegraphics[width=8.5cm]{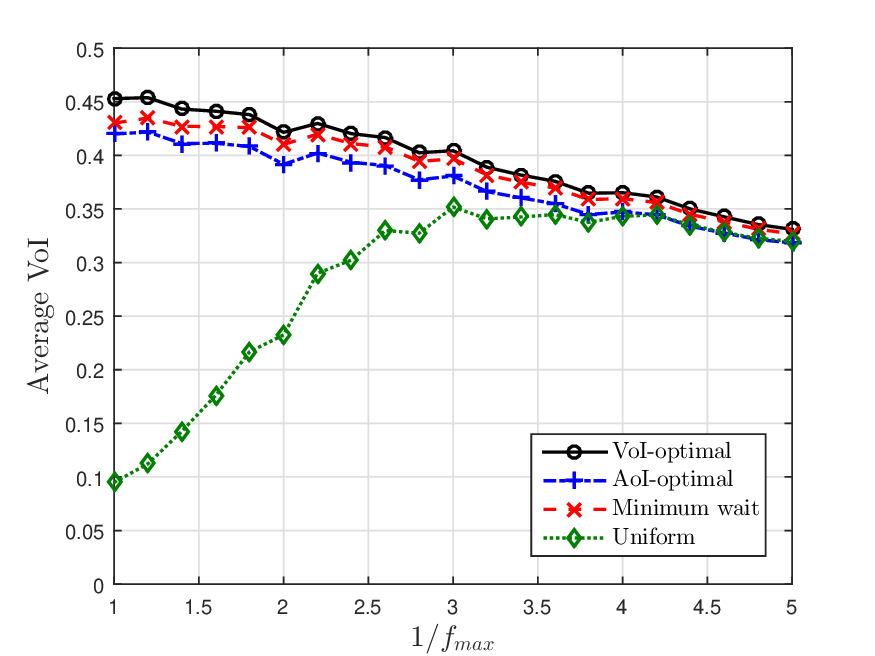}
\caption{Average VoI in the noisy OU process versus $1/f_{\max}$ for different sampling policies; transmission rate parameter $\mu=0.5$, correlation parameter $\kappa=0.1$ and the noise parameter $\sigma_n^2=0.1$.}
\label{fig:sampling voi vs Tmin }
\end{figure}

Figs.~\ref{fig:sampling voi vs mu } and~\ref{fig:sampling voi vs Tmin } illustrate the performance of the VoI under different sampling policies. In Theorem~\ref{theo:sampling policy}, we show that the optimal sampling policy to maximise the average VoI is a threshold-based policy. In addition to this VoI-optimal sampling policy, the following three policies are considered to evaluate the network-level information value.
\begin{itemize}
    \item AoI-optimal policy~\cite{QueueSYUpdatWait}: the optimal threshold-based solution to minimise the average AoI.
    \item Minimum wait policy: a new sample is generated once the previous sample is delivered under the sampling constraint, i.e., $z(S_i)=\max(0,\frac{1}{f_{\max}}-S_i)$ where $S_i$ represent the transmission time and $z$ represent the waiting time. It is the zero wait policy without the sampling constraint.
    \item Uniform sampling policy: the sampling intervals are constant $\Delta t$.
\end{itemize}
In the simulation, $\{S_i\}$ are generated randomly by a rate $\mu$ exponential random process.

Fig.~\ref{fig:sampling voi vs mu } shows the average VoI versus the transmission rate $\mu$ for different sampling policies without the sampling rate constraint. Here, the uniform sampling interval $\Delta t=\frac{1}{\mu}$. The average VoI increases with the transmission rate $\mu$ and the threshold-based sampling policy achieves the largest average VoI. Fig.~\ref{fig:sampling voi vs Tmin } shows the average VoI versus the minimum sampling interval $\frac{1}{f_{\max}}$ for different sampling policies. Here, the uniform sampling interval $\Delta t=\frac{1}{f_{\max}}$. As $\frac{1}{f_{\max}}$ goes larger, the constraint given in~\eqref{eq: problem formulation} will be an equality. This means that the sampling rate constraint is active, and the VoI-optimal policy will turn to the minimum wait sampling policy. These numerical results further illustrate that the threshold-based sampling policy achieves highest information value, and the VoI-optimal strategy is not equivalent to the AoI-optimal strategy.


\section{Conclusions}
In this paper, a mutual-information based value of information framework was formalised to characterise how valuable the status updates are for hidden Markov models. The notion of VoI was interpreted as the reduction in the uncertainty of the current unobserved status given that we have a dynamic sequence of noisy measurements. We took the noisy OU process as an example and derived closed-form VoI expressions. Moreover, the VoI was further explored in the context of the noisy OU model with different sampling times. The simplified VoI expressions were derived in high and low SNR regimes and statistical properties of VoI were obtained in M/M/1 queue model. The optimal sampling policy to maximise the average VoI was also investigated under the sampling rate constraint. Furthermore, numerical results are presented to verify the accuracy of our theoretical analysis. Compared with the traditional AoI metric, the proposed VoI framework can be used to describe the timeliness of the source data, how correlated the underlying random process is, and the noise in hidden Markov models. 

Even though we focus on the single transmitter and single receiver in this paper, one could imagine that there are ways to extend this framework to multiple transmitters and receivers. For example, we can consider the network where multiple sensing devices are deployed to monitor the same random process. The noisy observations of each device are correlated, but they are captured at different times and from different locations. The proposed framework can be used to evaluate the information value of the underlying random process given the noisy observations from different locations.

\begin{appendices} 
      \section{Proof of Proposition 1} 
      \label{appendix:voi}
      Since $(\bm{Y}^{\operatorname{T}},X_t)$ is multivariate normal distribution, the VoI defined in~\eqref{eq:general definition} can be written as~\cite{bookinformationtheory}
      \begin{equation}
      \label{eq:genernal voi with covariance}
          \begin{aligned}
              v(t)&=I(X_t;\bm{Y}^{\operatorname{T}}) \\
              &= h(X_t) + h(\bm{Y}^{\operatorname{T}}) - h(\bm{Y}^{\operatorname{T}},X_t)\\
              &=\frac{1}{2}\log \frac{{\Var[X_t]\det({\mathbf{\Sigma_Y} })}}{\det({\mathbf{\Sigma} _{\mathbf{Y},X_t}})}
          \end{aligned}
      \end{equation}
      where ${\mathbf{\Sigma_Y}}$ and $\mathbf{\Sigma} _{\mathbf{Y},X_t}$ are the covariance matrices of ${\bm{Y}}$ and $({\bm{Y}}^{\operatorname{T}},X_t)^{\operatorname{T}}$, respectively. 

Since ${\bm{X}}$ and ${\bm{N}}$ are independent, the covariance matrix ${\mathbf{\Sigma} _{\mathbf{Y}}}$ can be given as
\begin{equation}
\label{eq:cov y}
   {\mathbf{\Sigma_Y} } = {\mathbf{\Sigma_X} } + {\mathbf{\Sigma_N}}.
\end{equation}
Moreover, $\det(\mathbf{\Sigma} _{\mathbf{Y},X_t})$ in~\eqref{eq:genernal voi with covariance} can be obtained by the probability density function of $(\bm{Y}^{\operatorname{T}},X_t)$, and this density function can be further obtained by marginalising the joint density function of $(\bm{Y}^{\operatorname{T}},X_t,\bm{X}^{\operatorname{T}})$ over $\bm{X}^{\operatorname{T}}$. Hence, we have
\begin{equation}
\label{eq:cov y xt}
    \det({\mathbf{\Sigma} _{\mathbf{Y},X_t}}) =\Var[{X_t}|{X_{{t_n}}}]  \det\bigg({\mathbf{\Sigma_N} } + {\mathbf{\Sigma_X} } + \frac{{{\mathbf{\Sigma_X}}\bm{v}\bm{v}^{\operatorname{T}}{\mathbf{\Sigma_N} }}}{{\Var[{X_t}|{X_{{t_n}}}]}}\bigg)
\end{equation}
where vector $\bm{v} = [0, \cdots ,0,{e^{ - \kappa (t - {t_n})}}]^{\operatorname{T}}$. 

Substituting~\eqref{eq:cov y} and~\eqref{eq:cov y xt} into~\eqref{eq:genernal voi with covariance}, the VoI for the noisy OU process can be expressed as
\begin{multline}
    v(t)\\
    =  \frac{1}{2}\log \left(\frac{{\Var[{X_t}]}}{{\Var[{X_t}|{X_{{t_n}}}]}} \frac{{\det({\mathbf{\Sigma_N} } + {\mathbf{\Sigma_X} })}}{{\det({\mathbf{\Sigma_N}} + {\mathbf{\Sigma_X} } + \frac{{{\mathbf{\Sigma_X}}\bm{v}\bm{v}^{\operatorname{T}}{\mathbf{\Sigma_N}}}}{{\Var[{X_t}|{X_{{t_n}}}]}})}}\right).
\end{multline}
By utilising the matrix determinant lemma, the determinant in the denominator can be written as
\begin{multline}
    {\det\bigg({\mathbf{\Sigma_N}} + {\mathbf{\Sigma_X} } +\frac{{{\mathbf{\Sigma_X}}\bm{v}\bm{v}^{\operatorname{T}}{\mathbf{\Sigma_N}}}}{{\Var[{X_t}|{X_{{t_n}}}]}}\bigg)}\\
    =\bigg(1+\frac{{{\bm{v}^{\operatorname{T}}}{{({\mathbf{\Sigma}^{ - 1}_{\mathbf{X}} + \mathbf{\Sigma}^{ - 1}_{\mathbf{N}}})}^{-1}}\bm{v}}}{{\Var[{X_t}|{X_{{t_n}}}]}}\bigg) {\det({\mathbf{\Sigma_N} } + {\mathbf{\Sigma_X} })}.
\end{multline}
Therefore, the VoI expression can be further written as
\begin{multline}
    v(t) = \frac{1}{2}\log \bigg(\frac{{1 }}{{1 - {e^{ - 2\kappa (t - {t_n})}}}}\bigg) \\
    - \frac{1}{2}\log \bigg(1 + \frac{{{2\kappa\sigma_n^2 }}}{{\sigma^2 \left(e^{2\kappa (t-t_n)} - 1\right) }} \frac{\det(\mathbf{A}_{mm})}{\det(\mathbf{A})}\bigg)
\end{multline}
where $\mathbf{A} = \sigma_n^2\mathbf{\Sigma}^{ - 1}_{\mathbf{X}} + \mathbf{I}$.
\section{Proof of Lemma~\ref{lemma: induction}} 
      \label{appendix:induction}
      Mathematical induction is utilised to prove the statement $f_k$ for all natural numbers $1\le k\le m$. 
      
      First, for the base case, we have 
      \begin{equation}
    {f_1} = 1+{\frac{1}{\gamma}}{a_1},\quad {f_1} = 0 \cdot  1+ {a_1}{\frac{1}{\gamma}} +{{{\frac{1}{\gamma^2}}}}.
\end{equation}
\begin{equation}
    \begin{aligned}
    {f_2} &= ({\frac{1}{\gamma}}{a_1} + 1)({\frac{1}{\gamma}}{a_2} + 1) - {{\frac{1}{\gamma^2}}}b_1^2,\\
    {f_2} &=  1+ ({a_1} + {a_2}){\frac{1}{\gamma}} + ({a_1}{a_2} - b_1^2){{\frac{1}{\gamma^2}}}.
\end{aligned}
\end{equation}
It is easy to see that $f_1$ and $f_2$ are clearly true.

Next, we turn to the induction hypothesis. We assume that, for a particular $s$, the cases $k=s$ and $k=s+1$ hold. This means that 
\begin{equation}
\label{eq:assumption}
    \begin{aligned}
    &{f_s} = 1 + \sum\limits_{i = 1}^s {{a_i}{\frac{1}{\gamma}} + \bigg(\sum\limits_{1 \le i < j \le s} {{a_i}{a_j} - \sum\limits_{i = 1}^{s - 1} {b_i^2} } \bigg){{\frac{1}{\gamma^2}}} + O({{\frac{1}{\gamma^3}}})} \\
    &{f_{s + 1}}=\\
    &1 + \sum\limits_{i = 1}^{s + 1} {{a_i}{\frac{1}{\gamma}} + \bigg(\sum\limits_{1 \le i < j \le s + 1} {{a_i}{a_j} - \sum\limits_{i = 1}^s {b_i^2} } \bigg){{\frac{1}{\gamma^2}}} + O({{\frac{1}{\gamma^3}}})}.
\end{aligned}
\end{equation}
As matrix $\mathbf{A}$ is tridiagonal, the cofactor expansion can be used to calculate the determinant. When $k=s+2$, we have the following recurrence relation
\begin{equation}
\label{eq:recur high snr}
    {f_{s + 2}} = ({\frac{1}{\gamma}}{a_{s + 2}} + 1){f_{s + 1}} - {{\frac{1}{\gamma^2}}}b_{s + 1}^2{f_s}.
\end{equation}
Substituting~\eqref{eq:assumption} into~\eqref{eq:recur high snr}, we have
\begin{equation}
\begin{aligned}
    &{f_{s + 2}} = \bigg({a_{s + 2}}\sum\limits_{i = 1}^{s + 1} {{a_i}}  + \sum\limits_{1 \le i < j \le s + 1} {{a_i}{a_j} - \sum\limits_{i = 1}^s {b_i^2} } \bigg){{\frac{1}{\gamma^2}}} \\
    &+ \bigg({a_{s + 2}} + \sum\limits_{i = 1}^{s + 1} {{a_i}} \bigg){\frac{1}{\gamma}} + 1 + {{\frac{1}{\gamma^2}}}b_{s + 1}^2 + O({{\frac{1}{\gamma^3}}})\\
    =& 1 + \sum\limits_{i = 1}^{s + 2} {{a_i}{\frac{1}{\gamma}} + \bigg(\sum\limits_{1 \le i < j \le s + 2} {{a_i}{a_j} - \sum\limits_{i = 1}^{s + 1} {b_i^2} } \bigg){{\frac{1}{\gamma^2}}} + O({{\frac{1}{\gamma^3}}})}.
\end{aligned}
\end{equation}
This shows that the statement $f_{s+2}$ also holds true, establishing the inductive step.

Both base cases and inductive steps are proved to be true, therefore we can conclude that $f_k$ in~\eqref{eq:f_m} holds for every $k$ with $1\le k\le m$.
\section{Proof of Corollary~\ref{cor: poi low snr}} 
      \label{appendix:cor general low snr}
    Let the matrix
    \begin{multline}
        \mathbf{B}=\gamma \mathbf{A}=\frac{\sigma^2}{2\kappa}\mathbf{\Sigma}^{ - 1}_{\mathbf{X}} + \gamma\mathbf{I}\\
        =\left[ {\begin{array}{*{20}{c}}
{{a_1+\gamma}}&{{b_1}}&{}&{}&{}\\
{{b_1}}&{{a_2+\gamma}}&{{b_2}}&{}&{}\\
{}&{{b_2}}& \ddots & \ddots &{}\\
{}&{}& \ddots &{{a_{m - 1}+\gamma}}&{{b_{m - 1}}}\\
{}&{}&{}&{{b_{m - 1}}}&{{a_m+\gamma}}
\end{array}} \right].
    \end{multline}
    For simplicity, we denote the function $g(\cdot)$ as  determinant ratio which is given as
    \begin{equation}
        g(\gamma)=\frac{{\det ({\mathbf{A}_{mm}})}}{{\gamma \det (\mathbf{A})}} = \frac{{\det ({\mathbf{B}_{mm}})}}{{\det (\mathbf{B})}} = {\bm{e}^{\operatorname{T}}}{\mathbf{B}^{ - 1}}\bm{e},
    \end{equation}
where vector $\bm{e}=[0,\cdots,0,1]^{\operatorname{T}}$.

The series expansion of $g(\gamma)$ at $\gamma=0$ is used to simplify the VoI in the low SNR regime. When $\gamma=0$, we have
\begin{equation}
    g(0)={\bm{e}^{\operatorname{T}}}\bigg({\frac{\sigma^2}{2\kappa}\mathbf{\Sigma}^{ - 1}_{\mathbf{X}}\bigg)^{ - 1}}\bm{e}=1.
\end{equation}
The first-order derivative of $g(\gamma)$ is given as
\begin{multline}
    \frac{{\operatorname{d}\!g(\gamma )}}{{\operatorname{d}\!\gamma }} = {\bm{e}^{\operatorname{T}}}\frac{{\operatorname{d}\!{\mathbf{B}^{ - 1}}}}{{\operatorname{d}\!\gamma }}\bm{e} =  - {\bm{e}^{\operatorname{T}}}{\mathbf{B}^{ - 1}}\frac{{\operatorname{d}\!\mathbf{B}}}{{\operatorname{d}\!\gamma }}{\mathbf{B}^{ - 1}}\bm{e}\\
    =  - {\bm{e}^{\operatorname{T}}}{\mathbf{B}^{ - 1}}{\mathbf{B}^{ - 1}}\bm{e},
\end{multline}
then we have
\begin{multline}
    g'(0) = - {\bm{e}^{\operatorname{T}}}{\bigg(\frac{2\kappa}{\sigma^2}\mathbf{\Sigma_X}\bigg)\bigg(\frac{2\kappa}{\sigma^2}\mathbf{\Sigma_X}\bigg)}\bm{e}\\
    =-1 - \sum\limits_{j = 1}^{m - 1} {{e^{ - 2\kappa \sum\limits_{i = j}^{m - 1} {{T_i}} }}}.
\end{multline}
Therefore, the determinant ratio can be given as
\begin{equation}
\label{eq:poi low snr ratio}
    \frac{{\det ({\mathbf{A}_{mm}})}}{{\gamma \det (\mathbf{A})}}=1-\bigg(1+ \sum\limits_{j = 1}^{m - 1} {{e^{ - 2\kappa \sum\limits_{i = j}^{m - 1} {{T_i}} }}}\bigg)\gamma+O(\gamma^2).
\end{equation}
The result given in~\eqref{eq:poi appro. low snr} is obtained by substituting~\eqref{eq:poi low snr ratio} into~\eqref{eq:prop. 1 general voi}.
      
\section{Proof of the Determinant Calculation for Uniform Sampling} 
    \label{appendix:characteristic equation}
    When the sampling intervals are constant, the inverse covariance matrix of $\bm{X}$ in~\eqref{eq: Sigma_x inverse general} can be written as
\begin{equation}
\label{eq:matrix x^-1}
     \mathbf{\Sigma}^{ - 1}_{\mathbf{X}} = \frac{{2\kappa }}{{{\sigma ^2}(1 - {\rho ^2})}}\left[ {\begin{array}{*{20}{c}}
1&{ - \rho }&{}&{}&{}\\
{ - \rho }&{1 + {\rho ^2}}&{ - \rho }&{}&{}\\
{}&{ - \rho }& \ddots & \ddots &{}\\
{}&{}& \ddots &{1 + {\rho ^2}}&{ - \rho }\\
{}&{}&{}&{ - \rho }&1
\end{array}} \right].
\end{equation}
The matrix $\mathbf{A}$ in~\eqref{eq: poi A} is given by
\begin{equation}
    \mathbf{A} = \sigma_n^2\mathbf{\Sigma}^{ - 1}_{\mathbf{X}} +\mathbf{I}= \left[ {\begin{array}{*{20}{c}}
a&b&{}&{}&{}\\
b&c&b&{}&{}\\
{}&b& \ddots & \ddots &{}\\
{}&{}& \ddots &c&b\\
{}&{}&{}&b&a
\end{array}} \right],
\end{equation}
where
\begin{equation}
\label{eq:uni paprameter abc}
a = \frac{{1}}{{{\gamma}(1 - {\rho ^2})}} + 1,\quad b = \frac{{-\rho }}{{{\gamma}(1 - {\rho ^2})}},\quad c = \frac{{1 + {\rho ^2}}}{{{\gamma}(1 - {\rho ^2})}} + 1.
\end{equation}
Since the matrix $\mathbf{A}$ is tridiagonal, we are able to calculate its determinant~\cite{TridiagnalPossion}.

Let the $m$-dimensional circulant matrix $\bm{\eta}$ where
\begin{equation}
    {(\bm{\eta}) _{i,j}} = \left\{ {\begin{array}{*{20}{l}}
1&{i = m,j = 1}\\
1&{j = i + 1}\\
0&{\text{others}}
\end{array}} \right.
\end{equation}
\begin{equation}
\label{eq:det eta}
    \det (\bm{\eta} ) = {( - 1)^{m - 1}}.
\end{equation}
The product of matrix $\mathbf{A}$ and $\bm{\eta}$ can be partitioned into four blocks
\begin{equation}
\begin{aligned}
\label{eq:block matrix}
    \mathbf{A}\bm{\eta}  &= \left[ {\begin{array}{*{20}{l}}
0&\vline& a&b&0& \cdots &0\\
\hline
0&\vline& b&c&b&{}&{}\\
 \vdots &\vline& {}&b&c& \ddots &{}\\
0&\vline& {}&{}& \ddots & \ddots &b\\
b&\vline& {}&{}&{}&b&c\\
a&\vline& {}&{}&{}&{}&b
\end{array}} \right] \\
&= \left[ {\begin{array}{*{20}{l}}
\bm{{\eta_{11}}}&\vline& \bm{{\eta_{12}}}\\
\hline
\bm{{\eta_{21}}}&\vline& \bm{{\eta_{22}}}
\end{array}} \right].
\end{aligned}
\end{equation}
Taking the determinant of both side, then we have
\begin{equation}
\label{eq:detB}
    \det (\mathbf{A}) = \frac{{\det (\bm{\eta _{22}})\det(\bm{\eta _{11}} - \bm{\eta _{12}}\bm{\eta _{22}}^{ - 1}\bm{\eta _{21}})}}{{\det (\bm{\eta} )}}.
\end{equation}

Here,
\begin{equation}
\label{eq:det eta_22}
    \det (\bm{\eta_{22}})=b^{m-1}.
\end{equation}
As $\bm{\eta_{22}}$ is a tri-band Toeplitz matrix, the inverse matrix can be expressed by~\cite{UpperTriangular}
\begin{equation}
\label{eq:inverse eta_22}
    \bm{\eta _{22}}^{ - 1} = \left[ {\begin{array}{*{20}{l}}
{{J_1}}&{{J_2}}& \cdots &{{J_{m - 1}}}\\
{}&{{J_1}}& \ddots & \vdots \\
{}&{}& \ddots &{{J_2}}\\
{}&{}&{}&{{J_1}}
\end{array}} \right]
\end{equation}
where ${J_i}$ falls in the form of the following recurrence relation 
\begin{equation}
    {J_i} = - \frac{c}{b}{J_{i - 1}} -{J_{i - 2}}
\end{equation}
with ${J_1} = \frac{1}{b}$ and ${J_2} =  - \frac{c}{{{b^2}}}$. Substituting~\eqref{eq:block matrix}, ~\eqref{eq:det eta_22} and~\eqref{eq:inverse eta_22} into~\eqref{eq:detB}, we have
\begin{equation}
\label{eq:det B uniform}
    \det (\mathbf{A}) = {( - 1)^m}{b^{m - 1}}({a^2}{J_{m - 1}} + 2ab{J_{m - 2}} + {b^2}{J_{m - 3}}).
\end{equation}

The recurrence relation can be solved by the roots of the characteristic polynomial. The characteristic equation is given by
\begin{equation}
    {\lambda ^2} + \frac{c}{b}\lambda  + 1 = 0,
\end{equation}
and the eigenvalues are
\begin{equation}
    {\lambda _1} = \frac{{ - c + \sqrt {{c^2} - 4{b^2}} }}{{2b}},\quad {\lambda _2} = \frac{{ - c - \sqrt {{c^2} - 4{b^2}} }}{{2b}}.
\end{equation}
Thus, $J_i$ can be written as
\begin{equation}
    {J_i} = \frac{1}{{\sqrt {{c^2} - 4{b^2}} }}(\lambda _1^i - \lambda _2^i).
\end{equation}
Substituting $J_i$ into~\eqref{eq:det B uniform}, we obtain the following result.
\begin{multline}
\label{eq: uni A}
     \det (\mathbf{A}) = \frac{{{{( - 1)}^m}{b^{m - 1}}}}{{\sqrt {{c^2} - 4{b^2}} }}\bigg({a^2}(\lambda _1^{m - 1} - \lambda _2^{m - 1}) +\\
    2ab(\lambda _1^{m - 2} - \lambda _2^{m - 2}) + {b^2}(\lambda _1^{m - 3} - \lambda _2^{m - 3})\bigg),
\end{multline}
where
\begin{equation}
\label{eq:uni lambda12}
    {\lambda _1} = \frac{{ - c + \sqrt {{c^2} - 4{b^2}} }}{{2b}},\quad {\lambda _2} = \frac{{ - c - \sqrt {{c^2} - 4{b^2}} }}{{2b}}.
\end{equation}
Similarly, we also have
\begin{multline}
\label{eq: uni Amm}
    \det (\mathbf{A}_{mm}) = \frac{{{{( - 1)}^{m - 1}}{b^{m - 2}}}}{{\sqrt {{c^2} - 4{b^2}} }}\bigg(ac(\lambda _1^{m - 2} - \lambda _2^{m - 2}) + \\
    (ab + bc)(\lambda _1^{m - 3} - \lambda _2^{m - 3}) + {b^2}(\lambda _1^{m - 4} - \lambda _2^{m - 4})\bigg).
\end{multline}


\section{Proof of the Joint Density Function of Sampling Interval and System Time} 
      \label{appendix:mm1 joint PDF}
In the FCFS M/M/1 queueing system, the variables $S_n$, $W_{n+1}$ and $T_{n+1}$ are independent with each other, thus their joint PDF can be obtained by
\begin{equation}
    \begin{aligned}
        {f_{{S_n},W,T}}({s_n},w,t) &= {f_{{S_n}}}({s_n}){f_W}(w){f_T}(t) \\
    &= \lambda \mu (\mu  - \lambda ){e^{ - \lambda t - \mu w - (\mu  - \lambda )s_n}}.
    \end{aligned}
\end{equation}
The system time of the $(n+1)$th update $S_{n+1}$ can be expressed by
\begin{equation}
    {S_{n + 1}} = {({S_n} - {T_{n + 1}})^ + } + {W_{n + 1}}
\end{equation}
where the non-negative term represents the waiting time. Therefore, the joint PDF of $T_{n+1}$ and $S_{n+1}$ can be obtained by
\begin{multline}
    {f_{T,S}}(t,s) = \int_0^{ + \infty } {{f_{{S_n},W,T}}({s_n},s - {{({s_n} - t)}^ + },t)} \operatorname{d}\!{s_n}\\
 = \lambda \mu (\mu  - \lambda ){e^{ - \lambda t}}\bigg(\int_0^t {{e^{ - \mu s - (\mu  - \lambda ){s_n}}}} \operatorname{d}\!{s_n} \\
 + \int_t^{s + t} {{e^{ - \mu (s + {s_n} + t) - (\mu  - \lambda ){s_n}}}} \operatorname{d}\!{s_n}\bigg)\\
 = \lambda \mu {e^{ - \lambda t - \mu s}} - {\mu ^2}{e^{ - \mu (t + s)}} + \mu (\mu  - \lambda ){e^{ - \mu t - (\mu  - \lambda )s}}.
\end{multline}
\end{appendices} 
\bibliographystyle{IEEEtran}
\bibliography{IEEEabrv,papers}

\end{document}